\documentclass[fleqn,usenatbib]{mnras}

\usepackage{newtxtext,newtxmath}
\usepackage[T1]{fontenc}
\usepackage{graphicx}	
\usepackage{amsmath}	
\usepackage[usenames,dvipsnames]{xcolor}
\usepackage{hyperref}
\usepackage{ulem}
\usepackage{soul}

\usepackage{macros} 

\usepackage{cancel}
\usepackage{soul}

\title[Tidal evolution of anisotropic subhaloes]
      {The tidal evolution of anisotropic subhaloes: A new pathway to creating isotropic and cored satellites}

\author[B. T. Chiang, F. C. van den Bosch and H.-Y. Schive]
{Barry T. Chiang$^{1,2}$\thanks{E-mail: barry.chiang@yale.edu},
Frank C. van den Bosch$^{1}$, and Hsi-Yu Schive$^{2,3,4,5}$
\vspace*{8pt}
\\
$^{1}$Department of Astronomy, Yale University, New Haven, CT 06511, USA.\\
$^{2}$Institute of Astrophysics, National Taiwan University, Taipei 10617, Taiwan.\\
$^{3}$Department of Physics, National Taiwan University, Taipei 10617, Taiwan.\\
$^{4}$Center for Theoretical Physics, National Taiwan University, Taipei 10617, Taiwan.\\
$^{5}$Physics Division, National Center for Theoretical Sciences, Taipei 10617, Taiwan.}

\date{Accepted XXX. Received YYY; in original form ZZZ}

\pubyear{2025}

\begin{document}

\label{firstpage}
\pagerange{\pageref{firstpage}--\pageref{lastpage}}
\maketitle


\begin{abstract}
It is common practice, both in dynamical modelling and in idealised numerical simulations, to assume that galaxies and/or dark matter haloes are spherical and have isotropic velocity distributions, such that their distribution functions are ergodic. However, there is no good reason to assume that this assumption is accurate. In this paper we use idealised $N$-body simulations to study the tidal evolution of subhaloes that are anisotropic at infall. We show that the detailed velocity anisotropy has a large impact on the subhalo's mass loss rate. In particular, subhaloes that are radially anisotropic experience much more mass loss than their tangentially anisotropic counterparts. In fact, in the former case, the stripping of highly radial orbits can cause a rapid cusp-to-core transformation, without having to resort to any baryonic feedback processes. Once the tidal radius becomes comparable to the radius of the core thus formed, the subhalo is tidally disrupted. Subhaloes that at infall are tangentially anisotropic are far more resilient to tidal stripping, and are never disrupted when simulated with sufficient resolution. We show that the preferential stripping of more radial orbits, combined with re-virialisation post stripping, causes an isotropisation of the subhalo's velocity distributions. This implies that subhaloes that have experienced significant mass loss are expected to be close to isotropic, which may alleviate the mass-anisotropy degeneracies that hamper the dynamical modelling of Milky Way satellites. 
\end{abstract}



\section{Introduction}
\label{sec:introduction}

In the cold dark matter (CDM) paradigm of structure formation, the hierarchical nature of structure formation, with smaller haloes being accreted into bigger haloes, gives rise to a hierarchy of substructure, with subhaloes hosting sub-subhaloes, hosting sub-sub-subhaloes, etc. As these subhaloes orbit their hosts, they experience mass-loss due to the combined effects of dynamical friction, tidal stripping, and impulsive (tidal) heating \citep[e.g.,][]{MBW10}. Over time a quasi-equilibrium establishes with the accretion rate of new subhaloes being roughly balanced by the depletion of existing substructure, thus giving rise to a subhalo mass function that only evolves slowly in time, mainly just shifting in mass in tandem with the growth of the host mass itself \citep[e.g.,][]{Jiang.vdBosch.16}.

The abundance and demographics of dark matter substructure is a sensitive probe of the nature of dark matter. Numerous ongoing efforts therefore focus on constraining the amount of dark matter substructure using a variety of different techniques, including annihilation and decay signals of dark matter particles \citep[e.g.,][]{Strigari.etal.07, Pieri.etal.08,  Ando2019Galax768A}, gravitational lensing \citep[e.g.,][]{Vegetti.etal.14, Hezaveh.etal.16, Meneghetti.etal.20}, gaps in stellar streams  \citep[e.g.,][]{Carlberg.12, Erkal.etal.16, Bonaca.etal.19}, and the abundance of satellite galaxies \citep[e.g.,][]{Anderhalden.etal.13, Nadler.etal.21}.

In order for these efforts to yield reliable constraints on the nature of dark matter, it is prudent that we can make reliable predictions for the abundance of dark matter substructure for a given cosmological model. Given that the tidal evolution of substructure is highly non-linear, modelling all but the most idealised circumstances has proven analytically intractable, and most studies have therefore relied on cosmological $N$-body simulations \citep[e.g.,][]{Gao.etal.04, Diemand.etal.07, Springel.etal.08, Giocoli.etal.10} or on semi-analytical models that are calibrated against those simulation results \citep[e.g.,][]{Taylor.Babul.04, vandenBosch2005MNRAS3591029V, Zentner.etal.05, Pullen.etal.14, Jiang.vdBosch.16}.

In recent years, though, it has become evident that state-of-the-art cosmological simulations still suffer from `overmerging'\textemdash
the artificial disruption of dark matter subhaloes due to inadequate mass and/or force resolution \citep[e.g.,][]{Penarrubia2010MNRAS4061290P, vdBosch.Ogiya.18, vandenBosch2018MNRAS4743043V, Benson2022MNRAS5171398B}. The result is a significant, artificial suppression of the subhalo abundance, especially at small host-halo centric radii \citep[][]{Green2021MNRAS5034075G}.

This has prompted several studies based on high-resolution, idealised simulations to study the tidal evolution of subhaloes in a highly controlled setting that can avoid issues related to artificial disruption \citep[][]{Ogiya.etal.19, Green2019MNRAS4902091G, Errani2020MNRAS491, Benson2022MNRAS5171398B, Errani2021MNRAS50518E}. One of the key insights from these studies, corroborated by analytical studies, is that the inner cusp of an NFW subhalo is always preserved during tidal evolution, and physical tidal disruption can never occur \citep[][]{vandenBosch2018MNRAS4743043V, Errani2020MNRAS491, Errani2021MNRAS50518E, Amorisco2021arXiv211101148A, Stucker2023MNRAS5214432S}. Another key result is that subhaloes evolve along `tidal tracks'\footnote{Actually, this insight goes back to \citet{Hayashi2003ApJ584541H} and \citet{Penarrubia2008ApJ673}}, tight correlations between the bound mass fraction and parameters that characterise subhalo structure, such as peak circular velocity. This empirical fact implies that the density profile of a tidal remnant is fully specified by its initial (i.e., pre-infall) density profile and the fraction of its mass lost; it does not depend on how or when that mass was lost. Hence, all subhaloes with the same initial density profile evolve along the same tidal track, irrespective of the detailed host potential or the orbital parameters \citep{Penarrubia2010MNRAS4061290P}. Idealised simulations have been used to obtain empirical fitting functions for the tidal tracks, which have been independently corroborated by several studies \citep[e.g.][]{Green2019MNRAS4902091G, Errani2021MNRAS50518E, Amorisco2021arXiv211101148A, Benson2022MNRAS5171398B, Stucker2023MNRAS5214432S} and have found wide applications in semi-analytic modelling of subhhaloes and satellite galaxies \cite[e.g.][]{Yang2020MNRAS4983902Y, Jiang2021MNRAS502621J, Ahvazi2023arXiv230813599A, Green2022MNRAS5092624G, Du2024PhRvD110b3019D, Folsom2025MNRAS5362891F}. Furthermore, it has been suggested that, akin to the energy-lowering isothermal King model \citep{King1966AJ7164K}, the process of subhalo mass stripping can be accurately approximated as consecutive `shell peeling' in energy space \citep[e.g.][]{Drakos2017MNRAS4682345D, Drakos2020MNRAS494378D, Errani2022MNRAS5116001E}. As the precise `energy truncation' is uniquely determined by the host halo properties and subhalo orbital parameters, the tidal mass loss rate can be analytically predicted by knowing only the subhalo initial density profiles without regard to its internal velocity-distribution structure \citep{Drakos2022MNRAS516106D, Stucker2023MNRAS5214432S}.

However, there is one potential shortcoming with all these idealised simulations.  Without exception, they all assumed (sub)haloes to have isotropic velocity distributions, and thus an ergodic distribution function, $f(E)$, that only depends on energy. The main reason for this assumption is that it facilitates setting up the ICs. If the subhalo is both spherical and isotropic, its distribution function can be obtained easily and uniquely from its density distribution via the Eddington inversion formula \citep{Eddington1916MNRAS76572E, Binney&Tremaine2008}. 

However, high-resolution cosmological simulations have shown that, in a $\Lambda$CDM cosmology, the velocity structure of dark matter haloes clearly deviates from isotropy \citep[e.g.,][]{Diemand:2007qr, Navarro2010MNRAS40221N, Ludlow2011MNRAS4153895L, Klypin2016MNRAS4574340K, He2024arXiv240714827H}\footnote{Haloes are also found to exhibit velocity anisotropy in alternative dark matter paradigms, including warm dark matter \citep{Polisensky2015MNRAS4502172P} and self-interacting dark matter \citep[e.g.][]{Chua2021MNRAS5001531C, Shen2021MNRAS5064421S}.}. These studies find that haloes overall are radially anisotropic. More specifically, the velocity anisotropy,
\begin{equation}\label{betadef}
\beta(r) \equiv 1 - \frac{\sigma^2_\rmt(r)}{2\sigma^2_\rmr(r)}\,,
\end{equation}
relating the tangential and radial velocity dispersion profiles, $\sigma_\rmt(r)$ and $\sigma_\rmr(r)$, respectively, typically increases from $\beta \sim 0.2$ near the centre to $\beta \sim 0.3$ around the scale radius, where $\rmd\log\rho/\rmd\log r = -2$, and than falling again to $\sim 0.2$ near the virial radius. However, the halo-to-halo variance is large, especially in the halo outskirts, and there are some hints that the radial anisotropy is larger at higher redshifts \citep[][]{Klypin2016MNRAS4574340K}; there also appears to be clear halo-mass-dependence in $\beta(r)$ \citep{He2024arXiv240714827H}.
\begin{figure}
    \includegraphics[width=\linewidth]{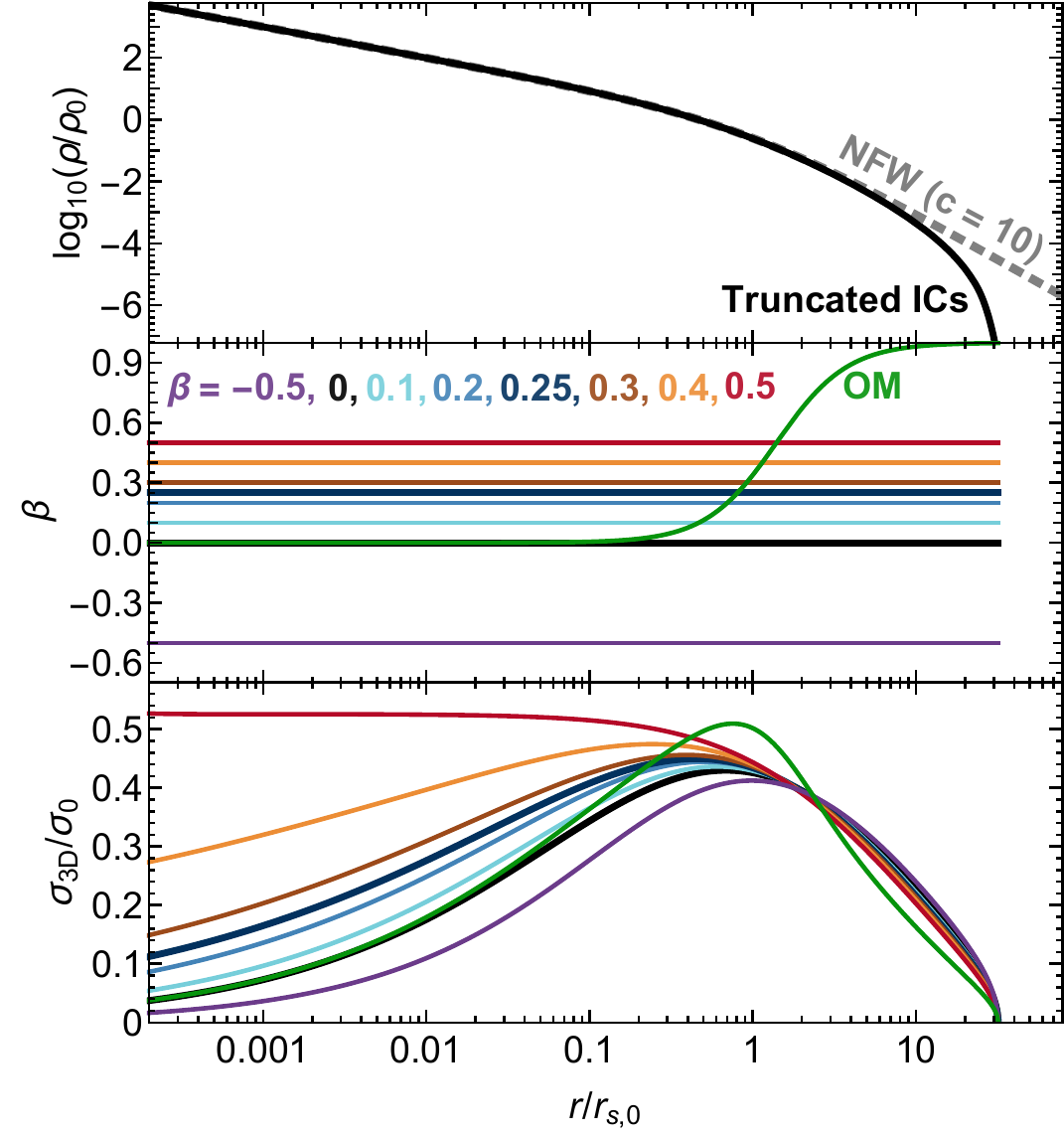}
	\caption{Initial density (top panel), velocity anisotropy (middle), and 3D velocity dispersion profiles (bottom) of all nine subhaloes discussed in the text (colour coded as indicated). The latter is normalised by $\sigma_0 \equiv \sqrt{G \msz/\rsz}$. All subhaloes have identical, phase-space-truncated density profiles, but differ in their velocity anisotropy profiles. Note how subhaloes that are more radially anisotropic have a higher velocity dispersion in their central regions, and how the Osipkov\textendash Merritt (OM) model transitions from being isotropic in the centre to maximally radially anisotropic ($\beta=1.0$) near the halo's truncation radius.}
	\label{fig:Analytical_IC}
\end{figure}

Given that anisotropy thus appears to be the rule rather than the exception, there is a clear need to understand how orbital anisotropy impacts the tidal evolution of dark matter subhaloes. This paper addresses this question using high-resolution, idealised simulations of subhaloes, with varying degrees of orbital anisotropy, orbiting in an analytical host-halo potential.  As we will demonstrate, (1) variation in the subhalo pre-infall velocity anisotropy can cause order-unity variations in the bound mass fraction within a Hubble time, (2) radially anisotropic subhaloes can undergo tidal-stripping-induced cusp-to-core transformation, and can experience complete, and physical, tidal disruption within a Hubble time, and (3) `tidal tracks' are not universal, but rather display a strong dependence on the initial (pre-infall) velocity anisotropy.
\begin{figure*}
	\includegraphics[width=\linewidth]{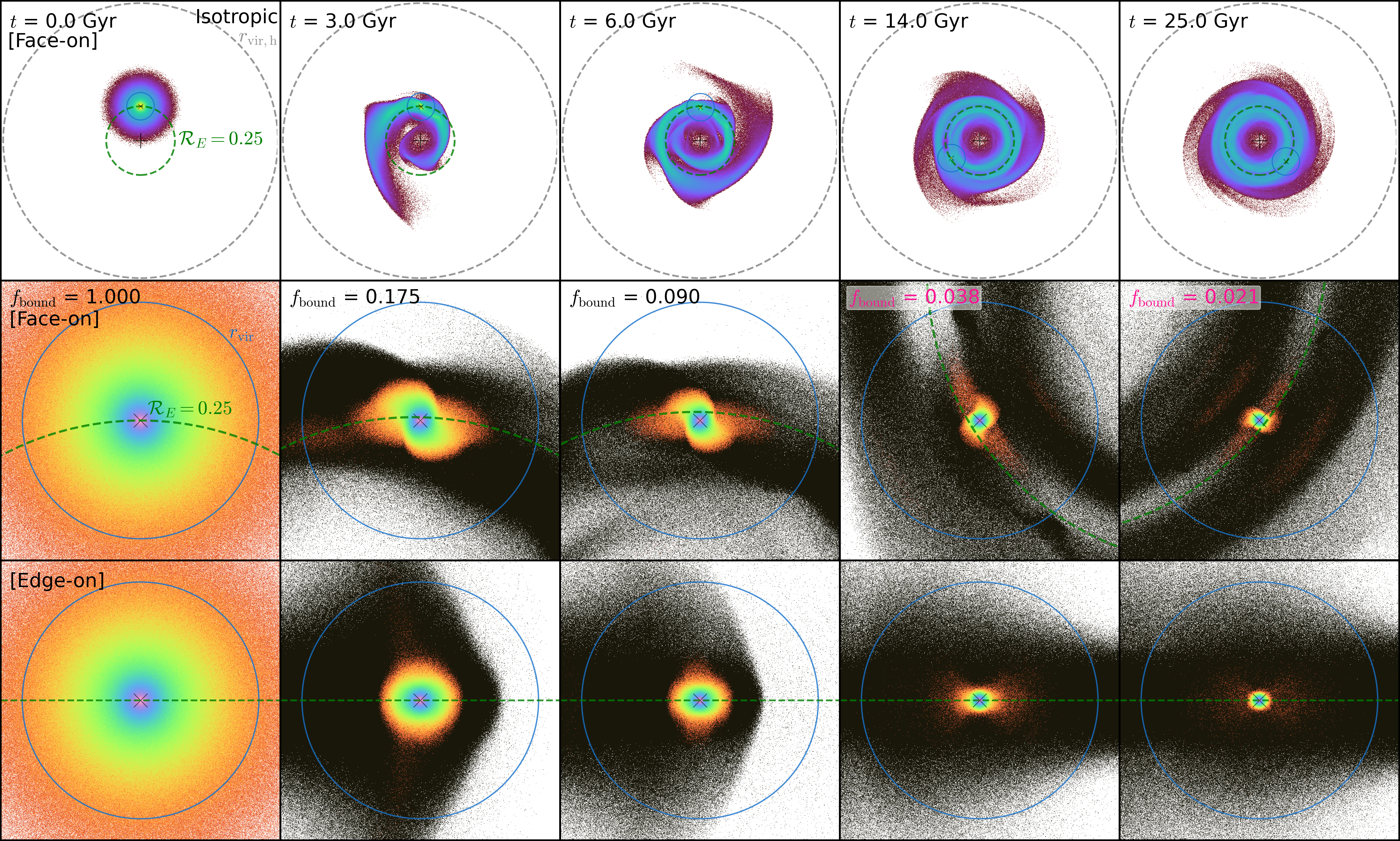}
	\caption{Tidal evolution of an isotropic subhalo resolved with $\Npar = 10^7$ particles on a circular orbit with $\bigRE = 0.25$. Different columns correspond to different epochs, as indicated.  \textit{Top row:} Face-on density projection of all subhalo particles, colour-coded by surface density. Grey and green, dashed circles indicate the virial radius of the host halo and the circular orbit of the subhalo, respectively. \textit{Middle and Bottom rows:} Face-on and edge-on zoom-in projections of the density of instantaneous bound (colour-coded) and unbound (black) particles. The blue, solid circle indicates the initial virial radius of the subhalo, anchored on its instantaneous centre of mass (marked by a cross), and the green, dashed curve marks the subhalo orbit. The bound mass fraction, $\fbound$, at each epoch is indicated in the panels in the middle row.}
	\label{fig:Isotropic_Combine}
\end{figure*}

The paper is organised as follows. In \S\ref{sec:Methodology}, we describe the initial conditions of our 
anisotropic, truncated NFW subhaloes and the numerical simulations used to tidally evolve these systems along a wide range of orbits. \S\ref{sec:results} describes the results, demonstrating a strong dependence of the tidal evolution of subhaloes on their pre-infall velocity anisotropy.  Next, we highlight and examine two new processes that we identified: tidal-stripping-driven core formation (\S\ref{ssec:coreformation}) and tidal isotropisation (\S\ref{ssec:Tidal_isotropisation}), and the implications thereof on alleviating the mass-anisotropy degeneracy for Milky Way (MW) satellites (\S\ref{ssec:Breaking_Mass_Ani_Degeneracy}). Finally, \S\ref{sec:Conclusions} summarises our conclusions.

\section{Methodology}
\label{sec:Methodology}

\subsection{Initial conditions and subhalo orbits}
\label{ssec:Simulation_Setup}

The main goal of this paper is to study how the tidal evolution of a subhalo depends on its velocity anisotropy. To that extent we construct initial conditions for nine subhaloes with identical density profiles, that only differ in their velocity anisotropy profiles. These subhaloes are then placed on different orbits in a static background potential (i.e., the `host halo') and integrated using a state-of-the-art $N$-body code.

Unless otherwise noted, all subhaloes are constructed with $\Npar = 10^7$ equal-mass particles sampled from spherically symmetric distribution functions that, before truncation, have an NFW density profile \citep{Navarro:1996gj}
\begin{align}\label{eqn:rho_NFW}
	\rho(r) = \rho_0\bigg(\frac{r}{\rsz}\bigg)^{-1}\bigg(1+\frac{r}{\rsz}\bigg)^{-2}\,,
\end{align}
with $\rsz$ the initial subhalo scale radius. The concentration of the subhalo is defined as $c \equiv \rvir/\rsz$, where the virial radius $\rvir$ is defined as the radius within which the average density is $\Delta_{\rm vir} = 97$ times the critical density \citep{Bryan1998ApJ49580B}. Throughout we restrict ourselves to subhaloes with $c=10$, which is a typical value for galaxy-sized haloes. The subhalo virial mass, $\msz$, denotes the total initial mass enclosed within $\rvir$ (in the absence of truncation). In order to avoid problems with the fact that the NFW profile extends to infinity, we truncate each subhalo using the methodology of \citet{Drakos2017MNRAS4682345D}, such that the total subhalo mass after truncation equals the untruncated subhalo virial mass.
\begin{figure*}
	\includegraphics[width=\linewidth]{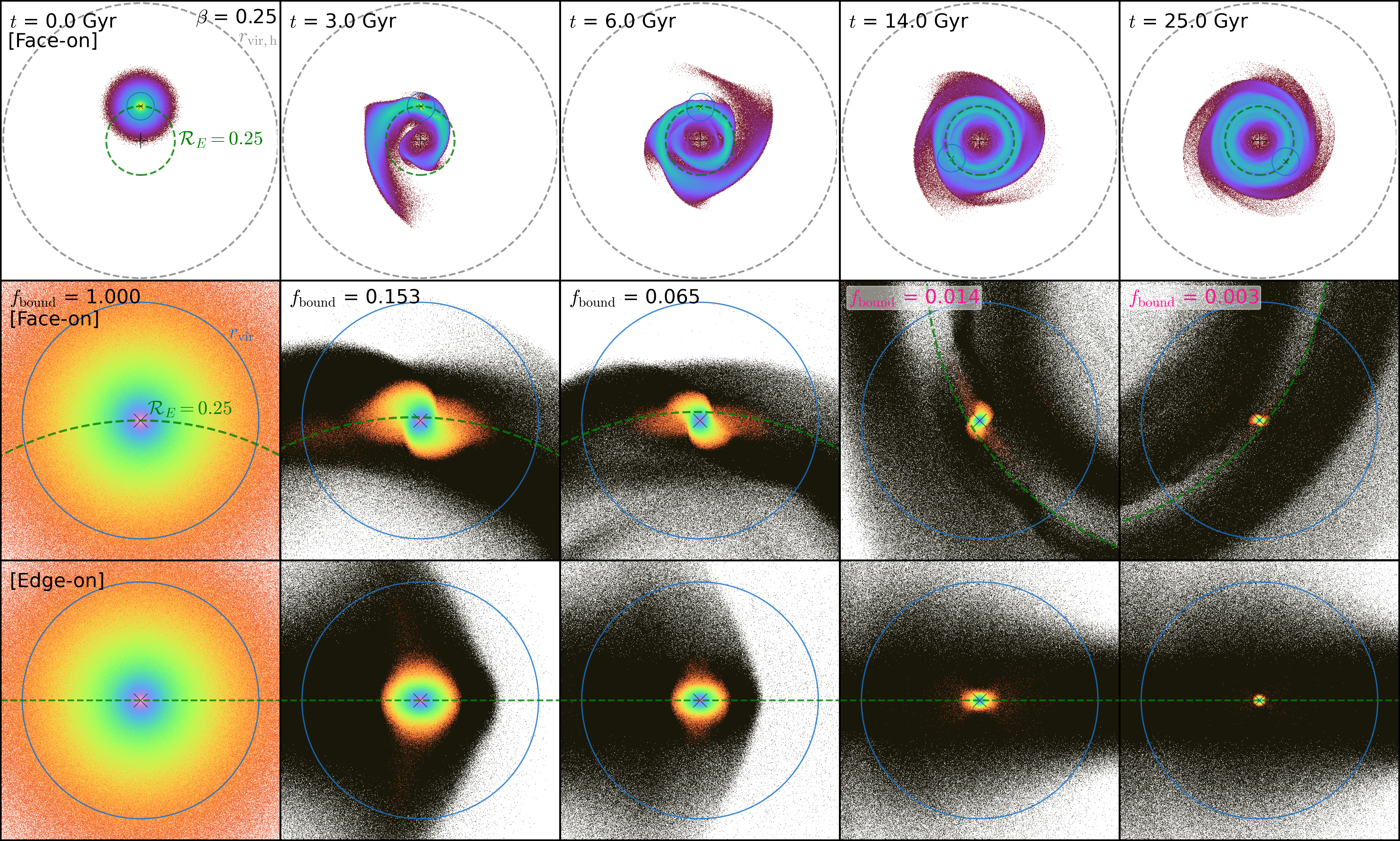}
	\caption{Same as \fref{fig:Isotropic_Combine} but for the radially anisotropic subhalo with $\beta(r) = +0.25$. Note that this subhalo undergoes significantly more mass stripping than its isotropic counterpart.}
	\label{fig:beta0p25_Combine}
\end{figure*}

Different subhaloes only differ in their velocity anisotropy profiles. Subhalo 1 has $\beta(r)=0$ and is thus isotropic throughout. Subhaloes 2 to 8 have a constant velocity anisotropy of $\beta(r) = -0.5, +0.1, +0.2, +0.25, +0.3, +0.4,$ and $+0.5$, respectively. Finally, subhalo 9 has an Osipkov\textendash Merritt (OM) distribution function \citep{Osipkov1979PAZh577O, Merritt1985AJ901027M} which implies that 
\begin{align}\label{eqn:OM_beta}
 \beta(r) = \frac{1}{1+\ra^2/r^2},
\end{align}
with $\ra$ the anisotropic radius. Hence, this system has an isotropic centre ($r \ll \ra$) and becomes more and more radially anisotropic at larger radii. Throughout this paper, we set $\ra = 1.4\rsz$. We also refer to subhaloes 1, 2, 3\textendash8, and 9 as the isotropic subhalo, the tangentially anisotropic subhalo, the radially anisotropic subhaloes, and the OM subhalo, respectively. Their (identical) density profiles, $\rho(r)$, anistropy profiles, $\beta(r)$, and 3D velocity dispersion profiles, $\sigma_{\rm 3D}(r)$ are shown in \fref{fig:Analytical_IC}.

The ICs for each of these subhaloes are generated using the new, open-source Python code \texttt{PIANISTpy} (\textbf{P}article \textbf{I}nitial conditions with \textbf{ANIS}o\textbf{T}ropy) described in Chiang et al. (in preparation). This code, which we specifically developed for this project, generates equilibrium initial conditions for spherical, double-power-law density profiles that are truncated in phase-space similar to the energy truncation methodology of \citet{Drakos2017MNRAS4682345D}, and that can have a wide range of different anisotropy profiles. By design, the self-consistent truncation of the system's distribution function does not alter the underlying velocity anisotropy profile $\beta(r)$, nor does it introduce any truncation-related instability, as for example in \citet[][]{Weinberg2022arXiv220906846W}. Importantly, the ICs require no further relaxation as the particle orbit sampling is exact, free of inherent `imperfections' that hamper approximate iterative schemes like the method of \citet{Schwarzschild1979ApJ232236S}.

Each subhalo is placed on an orbit inside the analytical potential of an NFW host halo with mass $M_\rmh = 10^3\msz$ and concentration $c_\rmh=5$. Using an analytical host halo implies that dynamical friction is ignored. However, this is not a concern, given that the ratio between host and subhalo mass is large enough that dynamical friction is negligible anyways. The orbits of the subhaloes are uniquely characterised by the orbital energy, $E_{\rm orb}$, and the orbital angular momentum, $L_{\rm orb}$. However, as is fairly common, we instead characterise the orbits using the dimensionless orbital radius
\begin{align}
\bigRE \equiv r_{\rm circ}(E_{\rm orb})/\rvirh,
\end{align}
and the orbital eccentricity
\begin{align}
e \equiv \frac{\rapo-\rperi}{\rapo + \rperi}.
\end{align}
Here $r_{\rm circ}(E)$ denotes the radius of a circular orbit of energy $E$, $\rvirh$ is the virial radius of the host halo, and $\rapo$ and $\rperi$ are the apo- and peri-centres of the orbit, which are the roots for $r$ of
\begin{align}\label{apoperi}
\frac{1}{r^2} + \frac{2[\Phi_\rmh(r)-E_{\rm orb}]}{L_{\rm orb}^2} = 0\,,
\end{align}
with $\Phi_\text{h}$ the gravitational potential of the host halo \citep[e.g.][]{vandenBosch1999ApJ51550V}. 

In this study, we vary $\bigRE$ over the range $[0.15,1.0]$ and mainly consider orbital eccentricities of $e=0$ (i.e., a circular orbit), $0.45$ and $0.9$. The orbital eccentricity is related to the orbital circularity, $\eta$, defined as the ratio of $L_{\rm orb}$, and the angular momentum $L_{\rm circ}(E_{\rm orb})$ corresponding to a circular orbit of energy $E_{\rm orb}$. For our host halo potential and $\bigRE=0.25$, eccentricities of $0$, $0.45$ and $0.9$ correspond to orbital circularities of $1.0$, $0.78$ and $0.17$, respectively. 

\subsection{Numerical simulations}
\label{ssec:numsim}

All simulations presented in this work are performed using the code $\gamer$ \citep{Schive:2017sdo}, which supports hybrid CPU/GPU parallelisation and adaptive mesh refinement (AMR) with octree data structure. We adopt a simulation box of dimensions $800\rsz\times800\rsz\times800\rsz$, covered by $128^3$ root grids and nine refinement levels, achieving a maximum spatial resolution of $\Delta x_\text{min} = 0.012\rsz$. For all our simulations we adopt the standard AMR strategy that further refines a cell if the total dark matter particle number therein exceeds four\footnote{Architecturally, $\gamer$ groups $8\times 8\times 8$ cells as a patch and performs refinement on a patch-by-patch basis. In addition, refinement levels of adjacent patches can vary by at most one level. This design differs from fully-threaded-tree AMR codes such as RAMSES \citep{Teyssier2002A&A385337T} that perform the refinement on a cell-by-cell basis. We have verified that our simulation results remain unchanged and numerically convergent under alternative mass+geometry-based AMR strategies.}. The large box size ensures that tidally stripped particles always remain sufficiently distant from simulation boundaries so as to minimise any possible numerical artifacts originating from the outflow boundary conditions. 

At each integration time step, the subhalo's instantaneous centre of mass (CoM) position and velocity are computed on-the-fly from the $5\%$ most bound particles, following the iterative procedure outlined in Section~2.3 of \citet{vdBosch.Ogiya.18}. In essence, during each iteration the subhalo self-potential is first computed from the instantaneous bound dark matter particles. Next, each individual dark matter particle's total energy, defined with respect to the subhalo's CoM, is updated and sorted, after which the new CoM position and velocity are determined using the $5\%$ most bound particles. This process is iterated until the changes in the CoM position and velocity are smaller than $10^{-4}\rvir$ and $10^{-4}V_\text{vir}$, respectively. The bound mass fraction is defined as
\begin{align}\label{eqn:f_bound}
\fbound(t) \equiv \frac{m_\rms(t)}{\msz}=\frac{N_{\rm bound}}{\Npar},
\end{align}
where $m_\rms(t)$ is the subhalo instantaneous bound mass at time $t$, $N_{\rm bound}$ denotes the number of instantaneous bound particles, and $\Npar = 10^7$ is the total number of simulated particles. 

We have carefully verified that the adopted numerical scheme is conservatively adequate to resolve the dynamical evolution of (initially) cuspy subhaloes and guarantees numerical convergence of the tidal evolution of our subhaloes down to $\fbound \leq 10^{-5}$. Additional numerical tests on subhalo stability, between $\gamer$ and \texttt{treecode}, and against different initial condition setups are presented in Appendix~\ref{app:Numerical_Convergence}. Further comprehensive convergence tests can be found in \citet{ChiangarXiv2025}.

Throughout we set $H_0 = 70 \kmsmpc$, yielding a Hubble time of $\tH \equiv H_0^{-1} = 13.97\Gyr$, and we adopt model units with $G = \rsz = \msz = 1$. With this choice, the initial virial velocity and crossing time of the subhalo are $\Vvir = c^{-1/2}$ and $t_{\rm cross} \equiv$ $\rvir/\Vvir = c^{3/2}$, respectively. Using that $t_{\rm cross}=2.0 \Gyr$, which follows from our definition of the virial overdensity, and that all subhaloes have $c=10$, we thus have that a time interval of $\Delta t=1$ (model units) corresponds to $63.4 \Myr$. For a subhalo on a circular orbit of radius $r_{\rm circ} = 0.25 r_{\rm vir,h}$ (our fiducial value), the corresponding orbital period is $\torb = 3.0\Gyr$, or $\torb = 47$ in model units.


\section{Results}
\label{sec:results}

\subsection{Tidal stripping of anisotropic subhaloes}
\label{ssec:Anisotropy_f_bound}

In order to probe the impact of internal velocity anisotropy on the tidal mass loss rate, we evolve all 9 subhaloes, which only differ in their initial velocity anisotropy, on circular orbits with $\bigRE = 0.25$, along which the tidal field strength remains time-invariant. Figs.~\ref{fig:Isotropic_Combine} and~\ref{fig:beta0p25_Combine} show snapshots of the tidal evolution of the isotropic subhalo and the radially anisotropic subhalo with $\beta = 0.25$, respectively. The top rows show the face-on (defined with respect to the orbital plane of the subhalo) density projection of all subhalo particles, colour-coded based on the total projected particle density. Green and gray dashed circles indicate the subhalo orbit and the virial radius of the host halo, respectively.  The middle and bottom rows of panels show face-on and edge-on zoom-ins on the subhalo. Particles that are bound are colour-coded by projected density, while particles that are (instantaneously) unbound are black. The blue circle indicates the virial radius of the initial subhalo, centred on the instantaneous CoM of the subhalo, and is shown for comparison. Note that the subhalo remnant drifts inward slightly over time. This is not due to dynamical friction against the host halo, which is absent since the latter is modeled as a static, analytical potential, but rather due to \textit{self}-friction against the tidally stripped material \citep[see][]{Miller2020MNRAS4954496M}.

\begin{figure}
	\includegraphics[width=\linewidth]{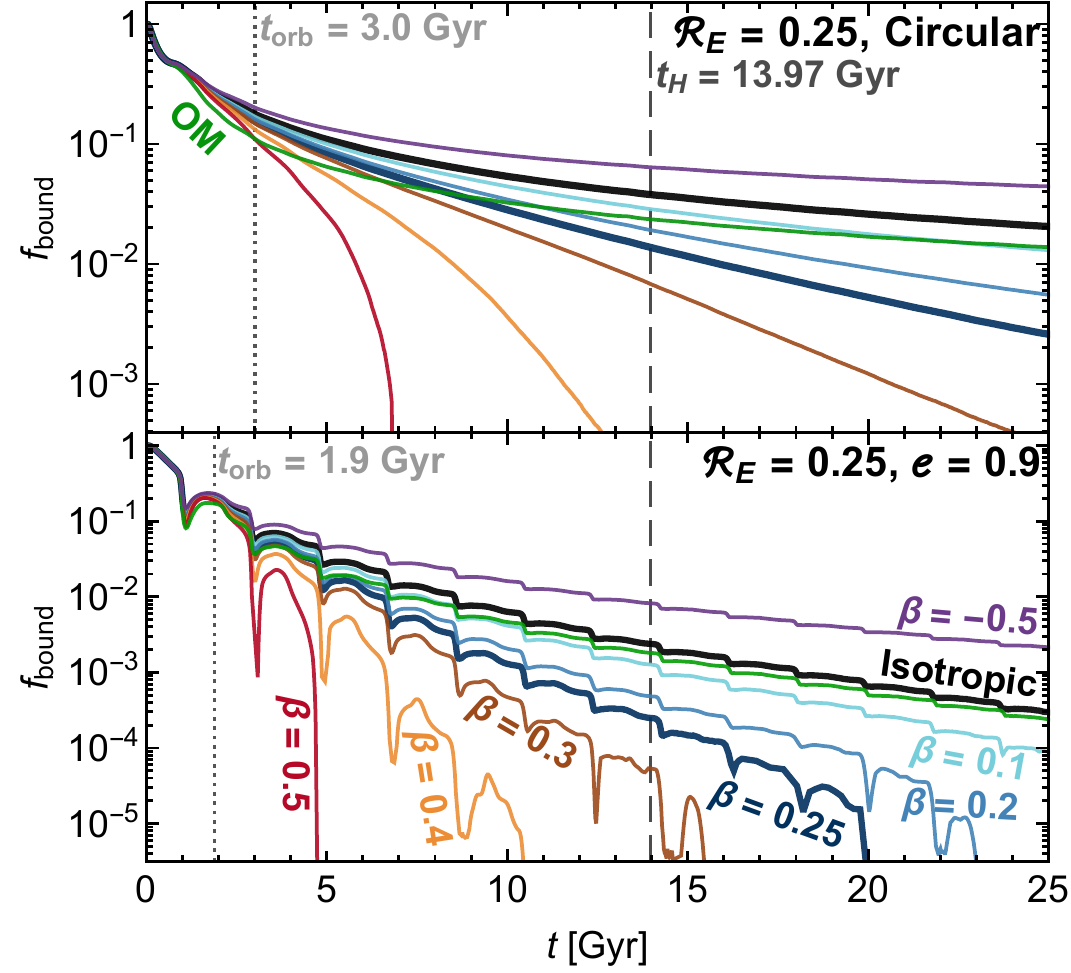}
	\caption{Evolution of the bound mass fraction of subhaloes along orbits with $\bigRE=0.25$ and eccentricities $e = 0$ (top) and $0.9$ (bottom), colour-coded by the initial anisotropy as indicated. The orbital time and Hubble time are indicated by vertical lines. Subhaloes that are more radially anisotropic experience more rapid mass loss, which in the most extreme cases can lead to complete physical disruption in under a Hubble time.}
	\label{fig:f_bound_Evolution}
\end{figure}

Note how over time the leading (inner) and trailing (outer) tidal tails morph into two ring-like structures on either side of the subhalo orbit, which itself appears as a gap-like feature. The main difference between the evolution of the isotropic (\fref{fig:Isotropic_Combine}) and radially anisotropic subhalo (\fref{fig:beta0p25_Combine}), as immediately evident upon inspection, is that the latter clearly experiences more pronounced mass stripping. By the last snapshot at $t=25\Gyr$, the bound mass fractions of these two subhaloes that start out with identical initial density profile already differ by an order of magnitude. This is the first indication that velocity anisotropy has a pronounced impact on tidal evolution.
\begin{figure*}
	\includegraphics[width=0.92\linewidth]{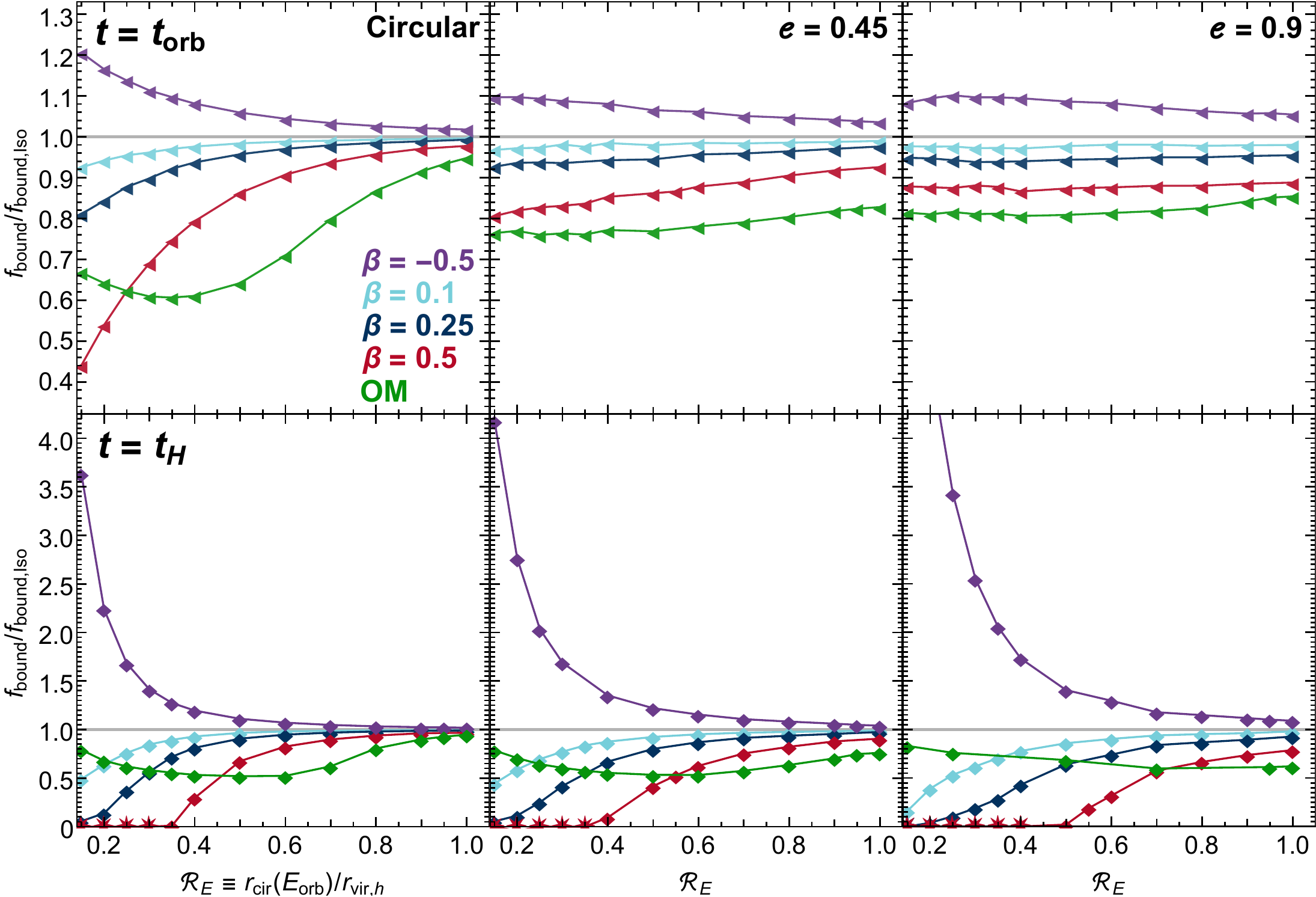}
	\caption{The ratio $\fbound/f_\text{bound,iso}$ of the bound mass fraction of an anisotropic subhalo to that of its isotropic counterpart, colour-coded as in \fref{fig:Analytical_IC}. Results are plotted as a function of the orbital energy, characterised by $\bigRE$, for three different orbital eccentricities (different columns, as indicated), and evaluated at two different epochs\textemdash after one orbital period, $\torb$ (top panels)  and after a Hubble time, $\tH$ (bottom panels). Star symbols indicate the cases where the subhalo is completely disrupted. Irrespective of the orbital parameters, more radially (tangentially) biased subhaloes are more susceptible to (resilient against) tidal stripping.} 
	\label{fig:f_bound_anisotropy}
\end{figure*}

This is further accentuated in \fref{fig:f_bound_Evolution}, which compares the evolution of the bound mass fraction, $\fbound(t)$, for all nine subhaloes evolved along two different orbits with $\bigRE = 0.25$: a circular orbit with $e=0$ (top panel) and a highly eccentric orbit with $e=0.9$ (bottom panel)\footnote{The dips in the $f_{\rm bound}(t)$ curves in the lower panel occur near peri-centric passages, when some constituent particles are temporarily unbound from the subhalo by the strong and impulsive tidal shocks but then become bound again during subsequent re-virialisation away from the peri-centre \citep[see][]{vdBosch.Ogiya.18}.}. As is evident, the amount of mass loss experienced by the subhaloes increases with increasing $\beta$; while the tangentially anisotropic subhalo with $\beta=-0.5$ experiences less mass loss than its isotropic counterpart, the radially anisotropic subhaloes experience an enhanced mass loss. By the end of a single orbit (indicated by the vertical, light-gray dotted line), the bound mass fractions of the various subhaloes already differ by up to a factor of $\sim 2$ compared to the standard `benchmark' isotropic subhalo. With the exception of the OM subhalo, this difference increases monotonically with time, and after a Hubble time (vertical dark-gray dashed line), the bound mass fraction can be different by orders of magnitude, especially along the eccentric orbit. In fact, the subhaloes with $\beta=0.5$ completely disrupt\footnote{Operationally, we define disruption as the epoch when we can no longer identify a bound remnant with $N_\text{bound} \leq 10$ (i.e. $\fbound \leq 10^{-6}$). We have verified that our results are insensitive to the exact choice of disruption threshold, given that $\fbound$ experiences a run-away decline after the onset of physical core disruption.} within under 7 (5) Gyr in the case of the circular (eccentric) orbit. For comparison, the tangentially anisotropic subhalo still has a bound mass fraction of $\sim 0.05$ ($\sim 0.01$) after a Hubble time.  In the case of the OM subhalo, the mass loss starts out even more extreme than for the $\beta=+0.5$ subhalo. However, after roughly one orbital period, when $f_{\rm bound}$ has dropped below $\sim 0.1$,  the bound mass fraction of the OM subhalo starts to asymptote towards that of the isotropic subhalo.  This behavior fits in with the overall trend, in that the outskirts of the OM subhalo are initially even more radially  anisotropic than that of the constant $\beta=+0.5$ subhalo (see \fref{fig:Analytical_IC}). However, once these outskirts have been stripped off, the remaining remnant is close to isotropic and its mass loss starts to resemble that of the isotropic subhalo.

To better highlight the dependence on the subhalo orbit, \fref{fig:f_bound_anisotropy} plots the ratio $\fbound/f_\text{bound,iso}$ of the bound mass fraction of anisotropic subhaloes to that of its isotropic counterpart, after having been tidally evolved for one orbital period $\torb$ (top row) or a Hubble time $\tH$ (bottom). Results are shown as a function of the orbital energy, as specified by $\bigRE$, for circular orbits (left-hand panel), mildly eccentric orbits with $e=0.45$ (middle panels), and highly eccentric orbits with $e=0.9$ (right-hand panels). Subhaloes that have experienced complete disruption are denoted by star symbols. As is evident, across the board the bound mass fractions are highly sensitive to the initial velocity anisotropy; radially (tangentially) biased subhaloes are significantly more (less) susceptible to tidal stripping compared to their isotropic counterparts. Typically the differences are larger for more bound and/or more eccentric orbits. This simply reflects that along such orbits the (maximum) tidal field strength is stronger. 

\begin{figure}
	\includegraphics[width=\linewidth]{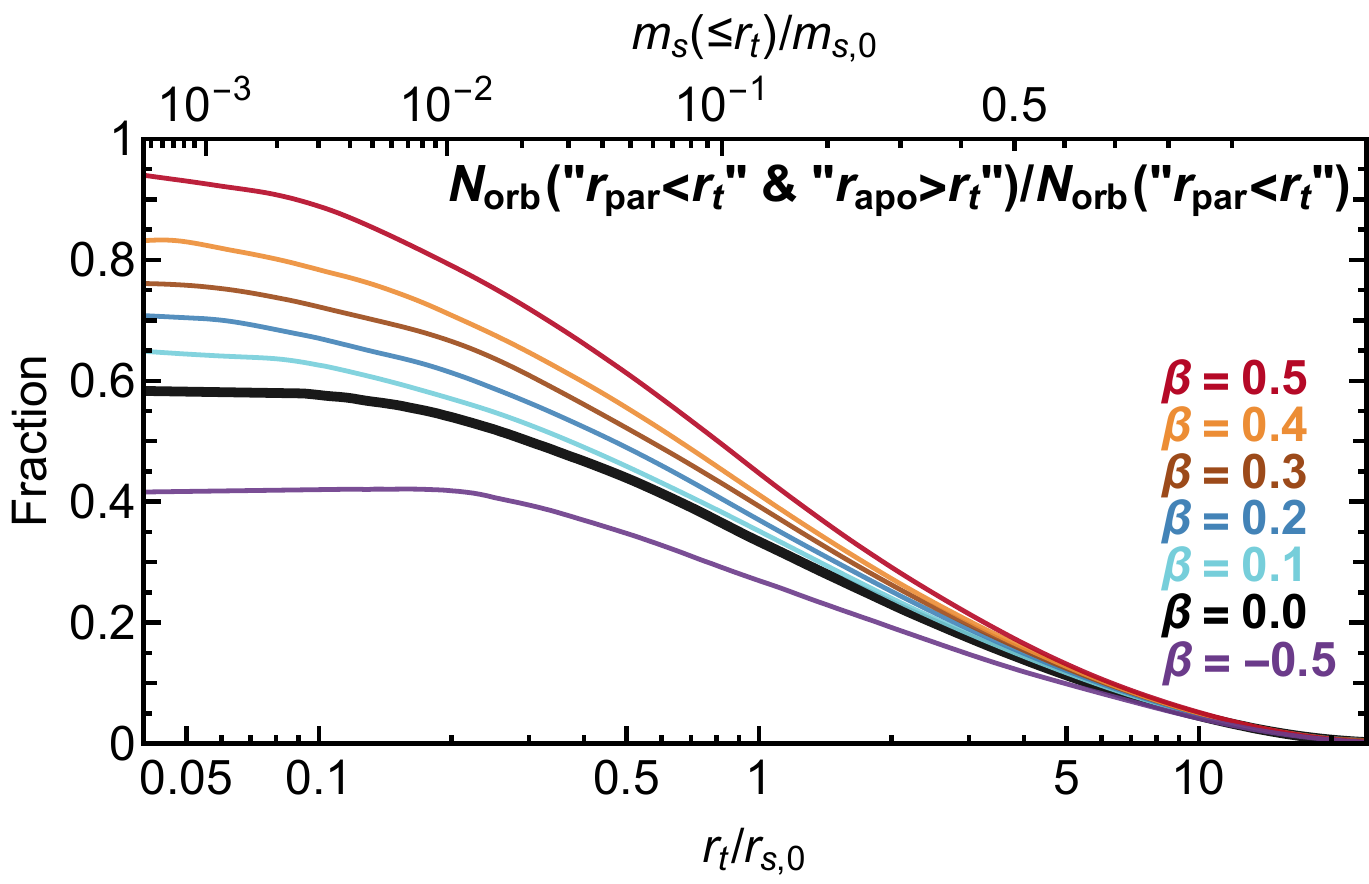}
	\caption{Fraction of constituent particle orbits, in the initial, unstripped subhalo, with instantaneous radius $r_\text{par} < r_\rmt$ that have an apo-centre radius $\rapo>r_\rmt$ as a function of a fictitious tidal radius, $r_\rmt$. Curves are colour-coded by the subhalo initial anisotropy as indicated. As discussed in the text, this fraction is indicative of the mass that is likely to be stripped within roughly one dynamical time. Since more radial orbits have larger apo-centres, subhaloes that are more radially anisotropic undergo more rapid stripping.}
	\label{fig:Orbit_Decomposition_I}
\end{figure}

One can understand the strong dependence of the tidal mass loss on the internal velocity anisotropy by considering their orbital composition. \fref{fig:Orbit_Decomposition_I} shows the fraction of particles, $f_\rmp$, with instantaneous radii $r_\text{par} < r_\rmt$ in the initial conditions that have apo-centric distances (computed using \eref{apoperi}) $r_{\rm apo}>r_\rmt$, as a function of some assumed value of the tidal radius, $r_\rmt$. Hence, these are the particles that are currently inside the tidal radius, but will orbit outside of $r_\rmt$, and thus be stripped, within less than one dynamical time. We can thus think of $f_\rmp$ as representative of the mass fraction that will be stripped off within about one dynamical time. Admittedly, this is crude, as it ignores the fact that prior stripping will have modified the density profile due to re-virialisation, and that $r_\rmt$ might well evolve in one dynamical time, but it suffices for the purpose of highlighting how the internal anisotropy impacts the stripping rate. Different colours correspond to the different initial anisotropies, as indicated. There is a clear trend of increasing $f_\rmp$ with increasing $\beta$. Particles in more radially anisotropic subhaloes are on more radial orbits, and thus, for a given orbital energy, have larger apo-centres, which in turn implies that they are more easily stripped. Note that this is a self-enhancing effect; once more mass is stripped off, the system re-virialises more strongly, which causes it to puff-up more strongly, which implies that the tidal radius moves in further, which promotes more stripping \citep[see][]{vandenBosch2018MNRAS4743043V}. This self-enhancement explains why the $f_{\rm bound}(t)$ curves for the various constant anisotropies in Fig.~\ref{fig:f_bound_Evolution} diverge from each other as time progresses.


\subsection{Non-universality of subhalo tidal tracks}
\label{ssec:Tidal_Tracks}

\begin{figure*}
	\includegraphics[width=0.99\linewidth]{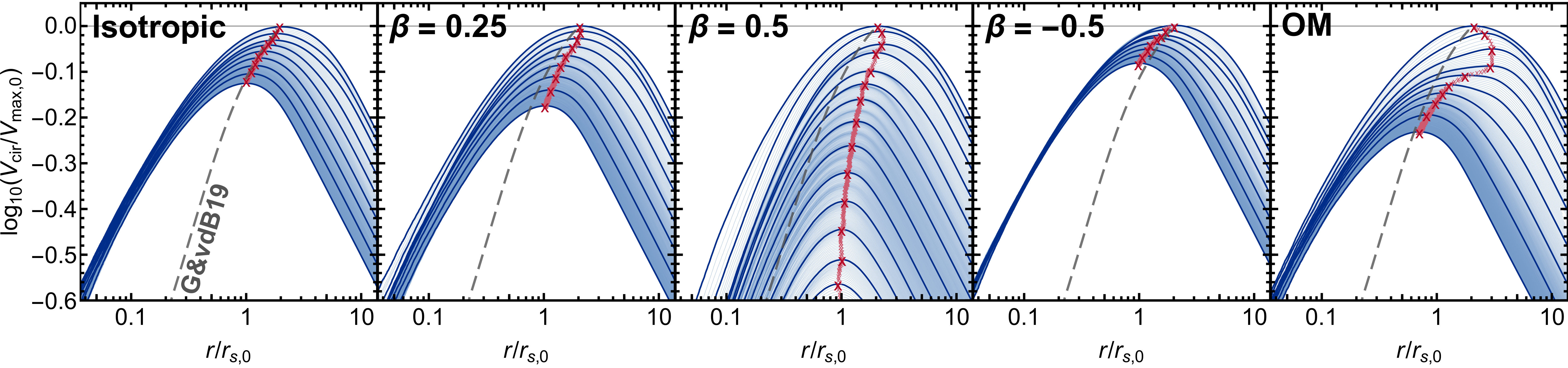}
	\caption{Circular velocity profiles of the bound remnants of subhaloes with different initial velocity anisotropy profiles (different panels, as indicated), tidally evolved along circular orbits with $\bigRE=0.5$.  Adjacent light-blue curves are offset temporally by $\Delta t = 0.1\Gyr$; dark-blue curves are shown at selected epochs to better highlight different evolution features. The profiles are normalised to the initial, maximum circular velocity $V_\text{max,0}$. The evolution of $(r_{\rm max}, V_{\rm max})$, marked by red crosses, traces out the tidal tracks. The gray-dashed curve is the best-fit tidal track for isotropic subhaloes obtained by \citet{Green2019MNRAS4902091G} and is shown for comparison. Note how  anisotropic subhaloes all yield distinct $(r_\text{max}, V_\text{max})$ evolutionary tracks that clearly deviate from the empirical, isotropic track.}
	\label{fig:Tidal_Tracks_Individual}
\end{figure*}

As discussed in \S\ref{sec:introduction}, one of the key insights that has emerged from idealised simulations of the tidal evolution of (isotropic) subhaloes is that they evolve along tidal tracks. This implies that the structure of a stripped subhalo only depends on its density profile at infall and its instantaneous bound mass fraction. However, as our analysis above has shown, the tidal evolution also depends strongly on the orbital anisotropy of the subhalo at infall, suggesting that the tidal tracks may not be universal.

\fref{fig:Tidal_Tracks_Individual} compares the evolution of subhalo circular velocity profiles, normalised to the maximum circular velocity at infall, $\Vmaxx$. The subhaloes are tidally evolved on circular orbits with $\bigRE = 0.5$, and different panels correspond to different initial anisotropies, as indicated. The red crosses indicate the tidal tracks, defined as the evolution in the parameter space of maximum circular velocity, $\Vmax$, and $\rmax$, the radius at which this maximum is achieved. Note how the tidal tracks for subhaloes with different initial anisotropy clearly differ from each other. The gray-dashed line is the best-fit tidal track of \citet{Green2019MNRAS4902091G}, obtained by fitting the results of idealised simulations for isotropic subhaloes, and is shown for comparison. As expected, the results for our isotropic subhalo (left-most panel) are in excellent agreement with this tidal track. However, this is not so for the anisotropic subhaloes, indicating that indeed the tidal tracks are not universal, but depend instead on the orbital velocity anisotropy of the subhalo at infall. 
\begin{figure*}
	\includegraphics[width=0.99\linewidth]{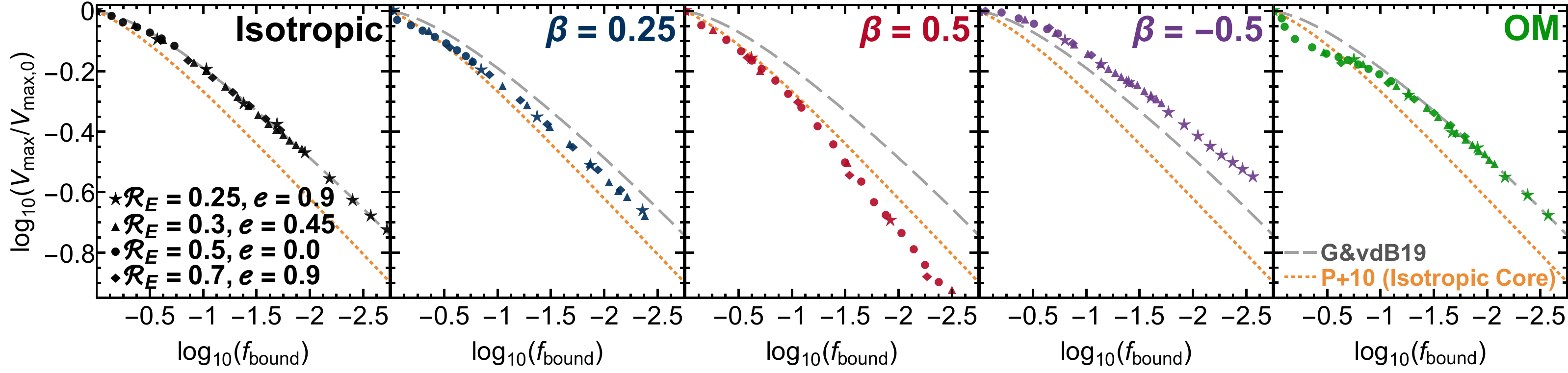}
	\caption{Tidal tracks, showing the evolution of the normalised peak circular velocity, $V_\text{max}/V_\text{max,0}$, against $\fbound$ for subhaloes with different initial velocity anisotropy, as indicated in the top-right corner of each panel. Results obtained along different orbits are indicated using different symbols, as indicated in the bottom-right corner of the leftmost panel. Data points of eccentric orbits are computed at apo-centre passages. Only isotropic subhaloes follow the well-established, isotropic tidal track \citep[e.g.][]{Green2019MNRAS4902091G} shown as a gray-dashed curve. The tidal tracks of radially (tangentially) biased subhaloes all lie below (above) this isotropic relation. For comparison, the orange-dotted curve shows the tidal track of an isotropic cored system obtained by \citet{Penarrubia2010MNRAS4061290P}.}
	\label{fig:Tidal_Tracks}
\end{figure*}

While tidal tracks may no longer be universal, we now examine whether at least the tidal evolution of subhaloes follows unique tidal tracks {\it for a given anisotropy profile}. For that purpose we evolve a setset of our subhaloes (the OM subhalo and those with constant $\beta=0$, $+0.25$, $+0.50$, and $-0.50$) along four different orbits  characterised by  $(\bigRE,e) = (0.25,0.9),(0.3,0.45),(0.5, 0.0),$~and~$(0.7,0.9)$. \fref{fig:Tidal_Tracks} plots the maximum circular velocity, normalised to that at infall, as a function of the instantaneous bound mass fraction, $\fbound$. Results for different orbits are indicates with different symbols, as indicated. For eccentric orbits we follow \citet{Penarrubia2010MNRAS4061290P}, and only show data points close to their apo-centres, where the subhalo is close to quasi-equilibrium. The gray-dashed is the tidal track for an isotropic, NFW subhalo from \citet{Green2019MNRAS4902091G}. As expected, the isotropic subhaloes follow this empirical isotropic tidal track independent of the adopted subhalo orbital parameters, consistent with previous studies \citep[e.g.][]{Hayashi2003ApJ584541H, Penarrubia2010MNRAS4061290P, Errani2021MNRAS50518E, Amorisco2021arXiv211101148A}. The anisotropic subhaloes, on the other hand, clearly deviate from this isotropic tidal track. However, they do follow equally narrow, evolutionary tracks that clearly seem to be independent of the subhalo orbital parameters. Hence, the tidal evolution of subhaloes does trace out tidal tracks, but they depend on the full density {\it and} velocity anisotropy profile of the subhalo at infall.

Note how the tidal tracks for the radially and tangentially anisotropic subhaloes fall, respectively, below and above that of the isotropic subhalo. As shown in \S\ref{ssec:coreformation}, the radially anisotropic subhalo with $\beta=+0.5$ undergoes a rapid, tidal cusp-to-core transformation, and both \citet{Penarrubia2010MNRAS4061290P} and \citet{Errani2023MNRAS519384E} have shown that cored subhaloes follow tidal tracks that fall below that of their isotropic counterparts. Interestingly, the tidal track of the $\beta=+0.5$ subhalo (middle panel) closely follows the tidal track for isotropic, cored profiles proposed by \citet{Penarrubia2010MNRAS4061290P} down to $\fbound \sim 0.1$, below which the two evolution tracks diverge. Note also how the tidal track for the OM subhalo initially drops significantly below the isotropic tidal track. However, once the radially biased outskirts have been stripped ($\fbound \lesssim 0.12$), leaving a close-to-isotropic cusp, the tidal evolution is `back on track', closely following that of the isotropic subhalo.
\begin{figure*}
	\includegraphics[width=0.99\linewidth]{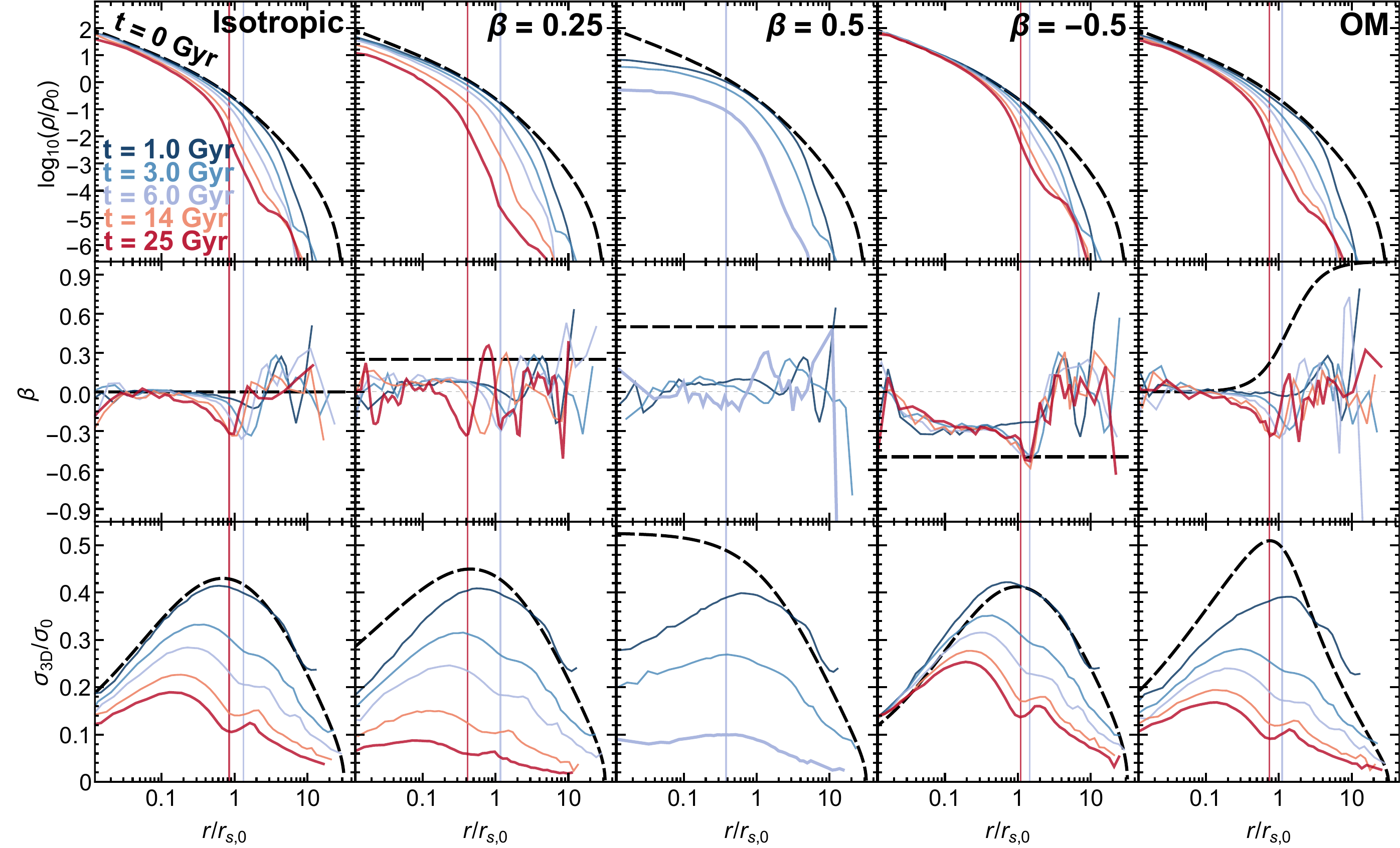}
	\caption{Tidal evolution of shell-averaged density (top row), anisotropy (middle), and 3D velocity dispersion profiles (bottom) of the bound remnants of subhaloes on circular orbits of $\bigRE = 0.25$ at various epochs between $t = 0$~Gyr (black dashed) and $t = 25$~Gyr (red solid), as indicated. Different columns correspond to subhaloes with different initial velocity anisotropy, as indicated in the top-right corner of the top panels. The light-blue and red vertical lines in each panel mark the instantaneous tidal radii, $\rtid$, at $t=6$~Gyr and $t=25$~Gyr, respectively. Note how the tidal remnants become more and more isotropic over time, independent of their initial anisotropy profile, how they reveal a `dip' in both $\beta(r)$ and $\sigma_\text{3D}(r)$ around $\rtid$, and how the $\beta=+0.5$ subhalo radidly transforms its initial cusp into a pronounced core, before undergoing complete disruption shortly after the tidal radius becomes comparable to the core radius at $t\simeq6.8$~Gyr (cf. Fig.~\ref{fig:f_bound_Evolution}).}
	\label{fig:Shell_Ave_Profiles}
\end{figure*}
%


\section{Tidal isotropisation and core formation}
\label{sec:Tidal_isotropisation}

Having established that the internal orbital anisotropy of dark matter subhaloes strongly impacts their tidal mass loss, we now shift focus to the tidal evolution of their internal structure. In particular, we highlight two intriguing phenomena; tidal core formation, which operates whenever the central region of a subhalo at infall is sufficiently radially anisotropic, and tidal isotropisation, which seems to be inevitable in all cases.

\fref{fig:Shell_Ave_Profiles} compares the time evolution of the density profile (top row), the velocity anisotropy profile (middle), and the 3D velocity dispersion (bottom) for the isotropic subhalo, the radially anisotropic subhaloes with $\beta=+0.25$ and $+0.5$, the tangentially anisotropic subhalo with $\beta=-0.5$, and the OM subhalo, as indicated. In each case, the orbit is the same circular orbit with $\bigRE = 0.25$ as in Figs.~\ref{fig:Isotropic_Combine} and~\ref{fig:beta0p25_Combine} and the top panel of Fig.~\ref{fig:f_bound_Evolution}. Different coloured lines correspond to different times, as indicated. For reference, we also mark the tidal radii, $\rtid$, at two of those epochs using vertical lines of matching colour. Here the tidal radius is defined as
\begin{align}\label{eqn:Tidal_Radius}
	\rtid = \left[\frac{G m(\rtid)}{\Omega^2 - \frac{\rmd^2\Phi_\rmh}{\rmd r^2}}\right]^{1/3},
\end{align}
\citep[e.g.,][]{King.62, Tollet.etal.17}, where the subhalo's angular velocity, $\Omega = v/r$, is defined with respect to the CoM of the host halo\footnote{See \citet{Tollet.etal.17} and \citet{vandenBosch2018MNRAS4743043V} for detailed discussions regarding different definitions of the tidal radius.}.

Evidently, density profiles exhibit distinctive evolution depending on the corresponding velocity anisotropy at infall. In the case of the tangentially anisotropic subhalo, the central density remains virtually unaffected for the entire duration of the simulation ($25\Gyr$), and the tides only affect (reduce) the density profiles in the outskirts. In the case of both the isotropic and OM subhaloes, the tides cause a slight reduction of the central densities, by about a factor of two. The reduction of the central density is even more pronounced (almost an order of magnitude) for the radially anisotropic subhalo with $\beta=+0.25$, while the $\beta=+0.5$ subhalo reveals a dramatic, and rapid, cusp-to-core transition; already after only $1\Gyr$ (half the subhalo crossing time) the inner density cusp has flattened noticeably, and by $t=6\Gyr$ it has developed a pronounced constant-density core in its centre. At this time, the instantaneous tidal radius, (indicated by the light-blue, vertical line) is comparable to the core radius, and the system undergoes complete disruption shortly hereafter.

The panels in the middle row of \fref{fig:Shell_Ave_Profiles} show the evolution in the velocity anisotropy profile, $\beta(r)$. The initially isotropic subhalo remains isotropic near the centre, but develops a pronounced `tangentially biased dip' around the instantaneous tidal radius, the `depth' of which first steadily deepens with time and saturates at $\beta\sim -0.3$. Note how a similar feature is also present in the other subhaloes, with the possible exception of the $\beta=+0.5$ case. As shown in \citet{Rozier.Errani.24}, such tangential dips are a natural outcome of tide-driven re-virialisation. The most striking aspect of the evolution of the anisotropy profiles, though, is that in all cases, they become more isotropic over time. In the case of the OM subhalo, the outskirts change from being strongly radially anisotropic to close to isotropic (with the exception of the tangentially anisotropic dip near $\rtid$) in under a Gyr, all the while preserving isotropy in its central region. The two radially anisotropic subhaloes that start out with $\beta=+0.25$ and $+0.50$, respectively, both have basically become completely isotropic, within the noisiness of the measurement and once again with the exception of the tangentially anisotropic dip, after less than $1\Gyr$ of tidal evolution. The tangentially anisotropic subhalo reveals the least amount of evolution, but even that system become close to isotropic in its central region, and again in a very short time. Hence, it appears that being exposed to a strong tidal field causes `tidal isotropisation'. We will investigate the cause of this effect in detail in \S\ref{ssec:Tidal_isotropisation} below.

Finally, for completeness, the lower panels of \fref{fig:Shell_Ave_Profiles} show the evolution in the (3D) velocity dispersion profiles. This evolution is qualitatively similar to that of the density profile; in the tangentially anisotropic case, the central region remains largely invariant, while the velocity dispersion in the outskirts becomes smaller as more and more mass is stripped off. The OM and isotropic subhaloes reveal a modest evolution (reduction) in the central velocity dispersion, while the velocity dispersion profile in the radially anisotropic cases drops precipitously over the entire radial range.
\begin{figure}
	\includegraphics[width=\linewidth]{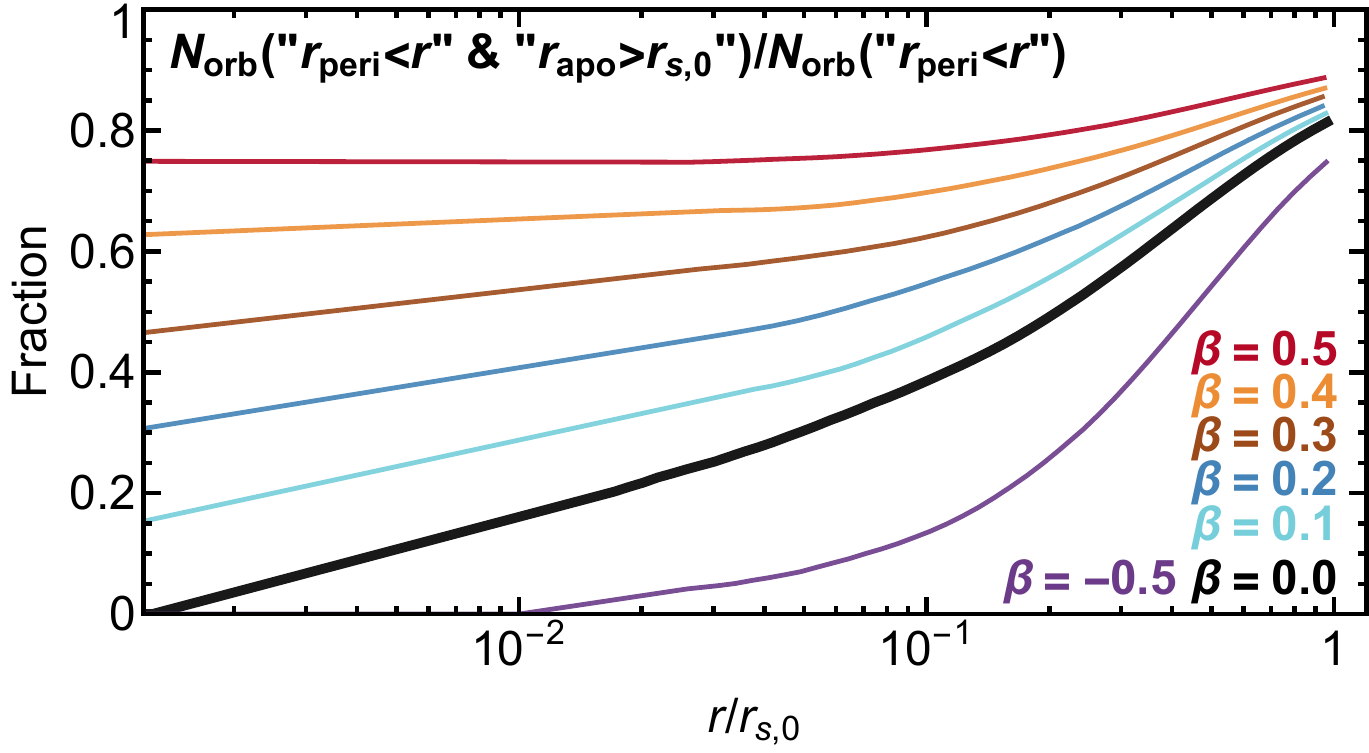}
	\caption{Fraction of constituent particle orbits with peri-centric distance $\rperi < r$ that have an apo-centre $\rapo > \rsz$, colour-coded by initial anisotropy as in \fref{fig:Orbit_Decomposition_I}. This fraction is indicative of the impact that tidal stripping with $r_\rmt=\rsz$ has on the density at radii $r < r_\rmt$. Since more radial orbits have smaller peri-centres, stripping of subhaloes that are more radially anisotropic affects the mass distribution down to smaller radii; this is how outside-in stripping of radially anisotropic subhaloes can directly affect the central density, promoting rapid core formation (cf. Figs.~\ref{fig:Shell_Ave_Profiles}~and~\ref{fig:Shell_Ave_Profile_Evolution_Ecc0p9}).}
	\label{fig:Orbit_Decomposition_II}
\end{figure}

\subsection{Tidal core formation}
\label{ssec:coreformation}

One of the key new insights of this study is that the tidal stripping of radially anisotropic subhaloes can cause a rapid cusp-to-core transition. In order to understand the physical origin of this tidal core-formation, it is once again useful to examine the subhalo's orbital composition. As we have seen in \S\ref{ssec:Anisotropy_f_bound}, since the orbits of more radially anisotropic orbits have larger apo-centres, a larger fraction of the subhalo's mass will be stripped off for a given tidal radius. \fref{fig:Orbit_Decomposition_II} shows the fraction of all particles in the initial conditions with a peri-centre $r_{\rm peri} < r$ that have an apo-centre $r_{\rm apo}>\rsz$, plotted as a function of $r$. Envision a tidal field such that the tidal radius $r_\rmt = \rsz$. On a circular orbit this means that, within roughly one dynamical time, all particles with $\rapo > r_\rmt = \rsz$ will be stripped off. \fref{fig:Orbit_Decomposition_II} then shows the fraction of those stripped particles that, prior to being stripped, contribute mass down to radii $<r$.   Since $\rperi \leq \rapo$, this fraction at $r=\rsz$ simply reflects the fraction of orbits with $\rapo > \rsz$, which, as we have already seen in \fref{fig:Orbit_Decomposition_I}, is larger in systems with larger $\beta$, which explains their higher mass-loss rates. More importantly, as is evident from \fref{fig:Orbit_Decomposition_II}, systems with larger $\beta$ also contribute more mass at small radii. In particular, in the case with $\beta=+0.5$, roughly 75 percent of all orbits that contribute mass below a scale of $0.01 \rsz$, will be stripped off once the tidal radius reaches down to $r_\rmt=\rsz$. For comparison, in the case with $\beta=0.0$, this fraction is only $\sim 15$ percent, while it is close to zero when $\beta=-0.5$. Hence, stripping of a more radially anisotropic subhalo directly impacts the subhalo further down towards its centre. Despite the fact that stripping only operates outside of the tidal radius, it can directly impact (lower) the central density, and the efficiency thereof depends on the peri-centres of the orbits that are being stripped. The further inwards the orbits that are being stripped penetrate, the more strongly the structural integrity of the subhalo's central density cusp is affected. 
\begin{figure}
	\includegraphics[width=\linewidth]{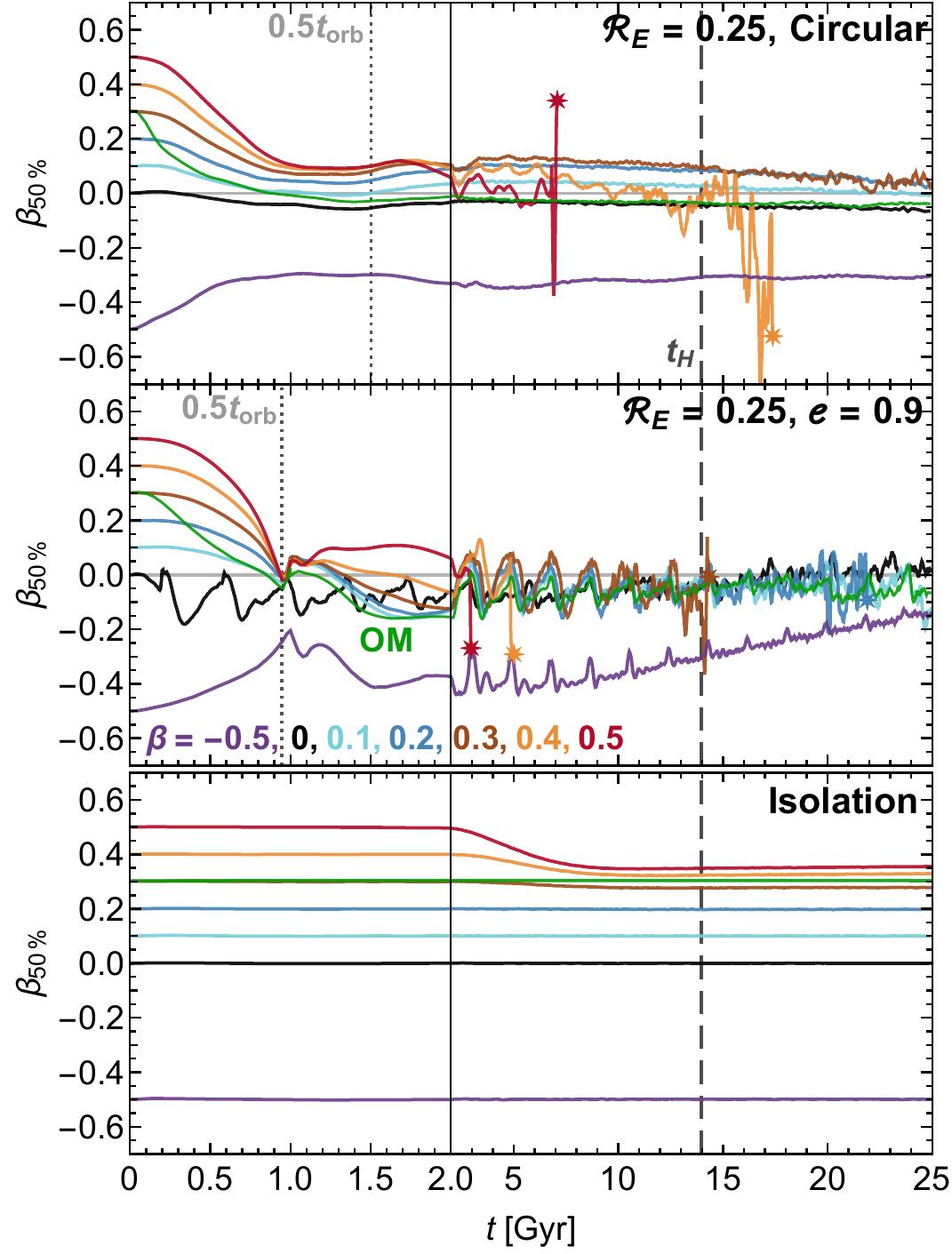}
	\caption{Evolution of $\beta_{50\%}$, defined as the velocity anisotropy of the instantaneous $50\%$ most bound particles, for subhaloes colour-coded by their initial velocity anisotropy (as indicated) and evolved on orbits with $(\bigRE,e) = (0.25,0.0)$ (top panel), $(0.25,0.9)$ (middle panel), or in isolation (bottom panel). Note that we have used two different scales for the time-axis in order to better highlight the rapid, leading-order isotropisation ($\beta_{50\%}\rightarrow 0$) within roughly half an orbital period (indicated by the vertical dotted line), and the much slower isotropisation due to tidal shocking that is most pronounced along more eccentric orbits. Note how prior to complete disruption (marked by stars), $\beta_{50\%}(t)$ undergoes large fluctuations. As is evident from the lower panel, systems with large radial anisotropy in their centres undergo a mild level of isotropisation in isolation due to the radial orbit instability. Importantly, this `spontaneous' isotropisation is clearly distinct from, and less pronounced than, the tidal isotropisation evident in the top and middle panels. See text for a detailed discussion.}
	\label{fig:Mass_Weighted_beta}
\end{figure}

It is this non-local aspect of tidal stripping, which is more pronounced for more radial orbits, that drives the rapid cusp-to-core transition in systems with sufficiently high radial anisotropy. This is aided by two processes that amplify the impact of the tidal stripping. First of all, as the system re-virialises in response to the mass loss, the system puffs up, which further contributes to lowering the (central) density. In addition, it `pushes' the apo-centres of the remaining particles to larger radii, which increases their probability of being stripped \citep[][]{Kampakoglou.Benson.07, Pullen.etal.14}. Secondly, in a central core all particles have similar orbital frequencies, while in a cusp more bound particles typically have higher frequencies. Orbits with higher frequencies are more likely to be adiabatically shielded from impulsive shocks that inject energy into the subhalo during peri-centric passages \citep[][]{Weinberg.94b, Gnedin.etal.99}. Since cusps are shielded from this energy injection, they typically are immune to tidal shocking, which explains why cusps are so resilient to tidal disruption. Cores on the other hand are typically not shielded, and will experience energy injection which, post re-virialisation, causes the core to expand during each peri-centric passage in the host halo \citep[e.g.,][]{Kampakoglou.Benson.07, Errani2020MNRAS491}. All these effects conspire to amplify the removal of mass from the central region, ultimately resulting in tidal core-formation. As tidal mass stripping continues, these cored subhalo bound remnants will undergo complete tidal disruption once the instantaneous tidal radius becomes comparable to the core radius \citep{Penarrubia2010MNRAS4061290P}.
\begin{figure*}
	\includegraphics[width=\linewidth]{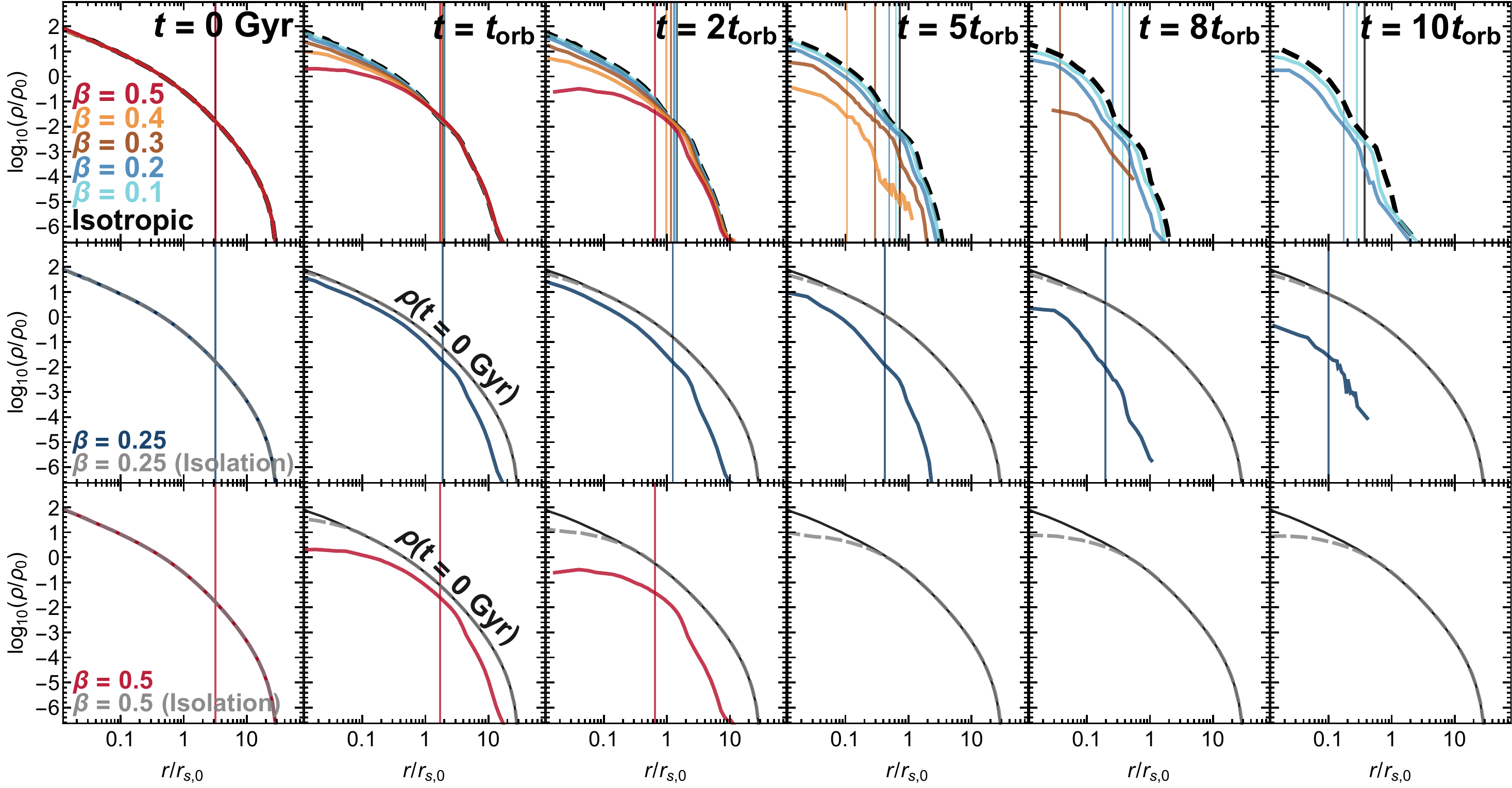}
	\caption{Shell-averaged density profiles of the bound remnants of subhaloes evolved on orbits with $\bigRE = 0.25$ and $e = 0.9$ (\fref{fig:f_bound_Evolution}), colour-coded by the initial anisotopy as in \fref{fig:Analytical_IC}. The leftmost to rightmost columns show time slices at $0$~Gyr and the subsequent apo-centre passages ($t_\text{orb} = 1.95$~Gyr), where the bound remnants are close to quasi-equilibrium. \textit{Top row:} More radially biased subhaloes undergo more rapid and pronounced core formation, due to preferential tidal stripping of constituent radial orbits (\fref{fig:Orbit_Decomposition_II}). Complete core disruption occurs shortly after the instantaneous tidal radii (vertical lines) become comparable to the respective core radii. The central core density and radius depend on both the initial anisotropy and $\fbound$. \textit{Middle and Bottom rows:} We additionally show profiles in isolation (gray dashed) and at $t =0$~Gyr (black solid). For the isolated subhaloes, the collective-effect-driven isotropisation (cf. \fref{fig:Mass_Weighted_beta}) at small radii can reduce the central density slope, but importantly on time and spatial scales distinct from the tide-driven core formation.}
	\label{fig:Shell_Ave_Profile_Evolution_Ecc0p9}
\end{figure*}

\subsection{Tidal isotropisation}
\label{ssec:Tidal_isotropisation}

As shown in Fig.~\ref{fig:Shell_Ave_Profiles}, all non-isotropic subhaloes experience rapid isotropisation of their orbital make-up once exposed to the tidal field. In order to better quantify this `tidal isotropisation', \fref{fig:Mass_Weighted_beta} plots the evolution of $\beta_{50\%}$, which we define as the average velocity anisotropy of the instantaneous $50\%$ most bound particles\footnote{This value of $50\%$ is chosen to capture a significant fraction of the instantaneous bound particles, while minimising noisy `contamination' from the nearly unbound orbits around $\rtid$, where the anisotropy profiles reveals a tangential dip (see Fig.~\ref{fig:Shell_Ave_Profiles}). We have verified, though, that the general conclusions drawn here are insensitive to this choice.}. Results are shown for all nine subhaloes, either in isolation (bottom panel), or along two different orbits characterised by $(\bigRE,e) = (0.25,0.0)$ and $(0.25,0.9)$ (top two panels, as indicated). Each subhalo is evolved for $25\Gyr$, or until complete tidal disruption, indicated by an asterisk. Note that the figure is split in two temporal sections, before and after $2\Gyr$, in order to better highlight the rapid evolution in the early stages.

The top panel shows the results for our fiducial circular orbit with $\bigRE=0.25$. Note that $\beta_{50\%}$ for the isotropic subhalo remains close to zero at all times, while that of the tangentially anisotropic orbit increases slightly from its initial value of $-0.5$ to $-0.3$, after which it remains nearly constant for the duration of the simulation. The situation for the OM and radially anisotropic subhaloes is very different. They all undergo rapid isotropisation, with $\beta_{50\%}$ dropping close to zero by $\sim 1\Gyr$, which is only one-third of the orbital period. The middle panel of \fref{fig:Mass_Weighted_beta} shows the temporal evolution of $\beta_{50\%}$ for our subhaloes along a highly eccentric orbit with $\bigRE=0.25$ and $e = 0.9$. Overall the results are similar to those for the circular orbit; once again the isotropic subhalo remains isotropic, while the OM and radially anisotropic subhaloes have $\beta_{50\%}$ drop rapidly to zero during the first Gyr of evolution. In the case of the tangentially anisotropic subhalo the evolution of $\beta_{50\%}$ is somewhat different. In particular, at late times the subhalo's velocity anisotropy becomes more and more isotropic over time, in a step-wise fashion. 

We emphasise that we find tidal isotropisation to be present along all orbits that we have explored
($0.15 \leq \bigRE \leq 1.0$; $0 \leq e \leq 0.9$), not just the two orbits depicted in Fig.~\ref{fig:Mass_Weighted_beta}. We have also investigated the evolution of our subhaloes in the energy–angular momentum space, measured in their rest frames, which clearly reveals isotropisation in that the distribution of angular momentum at a given energy becomes more homogeneous with time. In general, the rate of tidal isotropisation is faster and more pronounced if the maximum tidal strength along the orbit is stronger. Hence, at fixed $\bigRE$, isotropisation is faster along more radial orbits. Similarly at fixed orbital eccentricity, isotropisation is more rapid along more bound orbits (lower $\bigRE$).

What is the cause of this tidal isotropisation? Several mechanisms can be at play. First of all, as we have seen above, tides preferentially strip the more radial, low-angular momentum orbits, which have larger apo-centres than their high-angular momentum, more circular counterparts with the same orbital energy. Hence, stripping should naturally lead to a reduction of $\beta$. We believe that this is the main cause for the rapid isotropisation of the radially anisotropic subhaloes (but see below). Another mechanism that is likely to play a role is the re-virialisation that supersedes the mass stripping. During re-virialisation, particles change their energy and angular momentum in response to a time-varying potential (violent relaxation), and this tends to drive the system to isotropy, which corresponds to the lowest energy state \citep[][]{Doremus1971PhRvL26725D}. This is likely to be the main mechanism causing the slow, step-wise isotropisation of the tangentially anisotropic subhalo along the highly eccentric orbit; each `step' here corresponds to a peri-centric passage, during which the subhalo experiences an impulsive shock. Following each shock, the subhalo rapidly re-virialises, after which it is left with a value of $\beta_{50\%}$ that is somewhat closer to isotropy. An impulsive shock gives each particle in the subhalo a velocity-impulse, $\Delta \bv(\bx)$, which changes both the orbital energy and angular momentum of the particle. During the subsequent re-virialisation, both the energy and angular momentum of the particles once again undergo changes due to the time-varying potential. The net outcome of this repeated mixing in energy-angular momentum space will be to drive the system to a more isotropic state. 
\begin{figure*}
	\includegraphics[width=\linewidth]{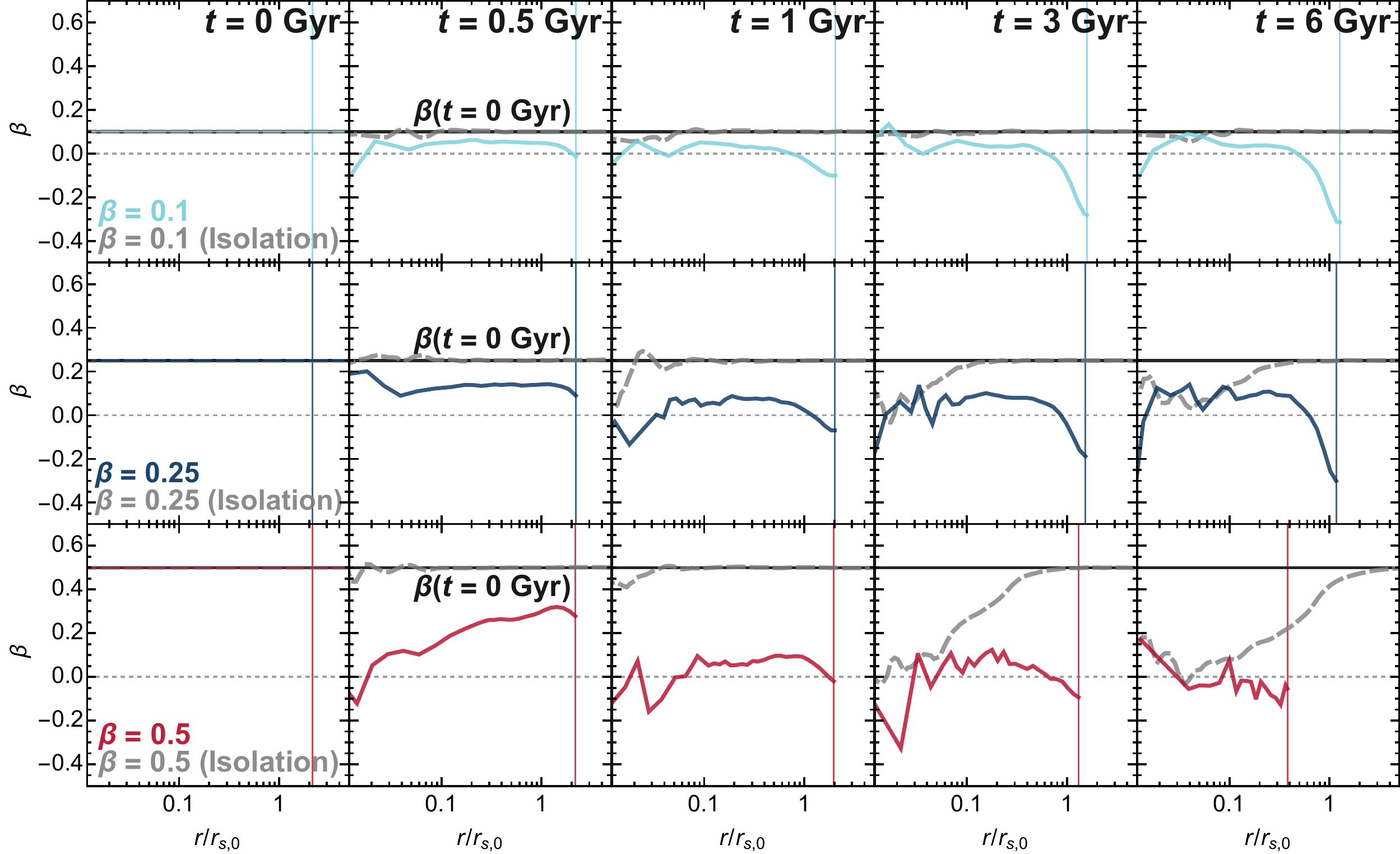}
	\caption{Shell-averaged velocity anisotropy profiles of the bound remnants of subhaloes evolved on circular orbits with $\bigRE = 0.25$ (cyan, dark blue, and red for $\beta = 0.1, 0.25,$ and $0.5$) or in isolation (gray dashed), as compared to the initial profile at $t =0$~Gyr (black solid). Results in the cases with tidal field are only shown out to the tidal radius, indicated with a vertical, coloured line. Although the systems with large, radial anisotropy ($\beta=0.25$  and $\beta=0.5$) experience isotropisation driven by the radial orbit instability, the isotropisation in the presence of tides is faster and more pronounced. See text for a detailed discussion.}
	\label{fig:beta_Evolution_Cir}
\end{figure*}

Another process that could potentially cause isotropisation is numerical, discreteness-driven collisionality. This can be tested by running the haloes in isolation (i.e., without the presence of an external tidal field). The lower panel of Fig.~\ref{fig:Mass_Weighted_beta}, though, shows that this is not a concern. For all nine subhaloes, $\beta_{50\%}$, remains perfectly invariant for the first $2\Gyr$, which is very different from the behavior seen in the presence of tides. In fact, with the exception of the subhaloes with $\beta \gta 0.3$, the value of $\beta_{50\%}$ remains invariant for the entire duration ($25\Gyr$) of the simulation. However, the most radially anistropic subhaloes with an initial $\beta \gta 0.3$ experience some `spontaneous' isotropisation in the time-interval from $\sim 3\Gyr$ to $8 \Gyr$. This is most pronounced for the $\beta = +0.5$ subhalo, which has $\beta_{50\%}$ dropping from $0.5$ to $0.36$, after which it stabilises. This spontaneous isotropisation does not originate from collisionality, though, but rather is caused by the radial orbit instability (ROI), which is evident from the steady development of prolate, bar-like isodensity surfaces in the innermost region of these systems on temporal and spatial scales that are consistent with those on which spontaneous isotropisation occurs. It is well known that systems that are strongly radially anisotropic in their centre are unstable to the ROI \citep[][]{Antonov1973, Henon1973, Polyachenko1981, Barnes1986}, and as hinted at in \citet{Trenti.Bertin.06}, the ROI has a tendency to drive the system towards a more isotropic state and with a reduced central density.

This poses the question whether the tidal isotropisation and perhaps even the tidal core formation seen in our simulations might perhaps be a manifestation of the ROI, rather than a pure outcome of tidal stripping as argued above. In order to address this, the top panels of Fig.~\ref{fig:Shell_Ave_Profile_Evolution_Ecc0p9} shows the density profiles of subhaloes evolved on orbits with eccentricity of $0.9$ and $\bigRE = 0.25$ (as in the bottom panel of \fref{fig:f_bound_Evolution}), colour-coded by their initial anisotopy, as indicated.  Different columns correspond to different apo-centric passages, and vertical lines in matching colour indicate the instantaneous tidal radii, computed using \eref{eqn:Tidal_Radius}.  For the sake of clarify, we only show results for subhaloes with a constant, initial anisotropy ranging from $\beta=0$ to $\beta=0.5$. As is evident, systems with a larger $\beta$ undergo more prompt and more pronounced core formation. Complete core disruption occurs whenever the core radius becomes comparable to the instantaneous tidal radius, after which the corresponding curve is no longer present in the subsequent columns. In order to contrast this to the ROI-driven evolution in the absence of tides, the lower two rows of panels compare the evolution of the $\beta=+0.25$ (middle row) and $\beta=+0.50$ (bottom row) systems in isolation (grey-dashed curves) to that in the presence of the tidal field (coloured curves), once again along an orbit with eccentricity of $0.9$ and $\bigRE = 0.25$.  As is apparent, the ROI also seems to cause a cusp-to-core transformation, at least in the $\beta=+0.5$ case, but the timescale of core formation is much longer. Similarly, \fref{fig:beta_Evolution_Cir} compares the evolution of the $\beta(r)$ profiles for subhaloes evolved on circular orbits with $\bigRE = 0.25$ (colour-coded) to those evolved in isolation (gray dashed), for three different values of the initial anisotropy. In the case with $\beta=+0.1$ (top panels), the system in isolation show no sign of evolution; the system is stable against the ROI, at least over the timescale of the simulation. In the cases with $\beta=+0.25$ and $\beta=+0.50$ the system in isolation does undergo ROI-driven isotropisation in its central region. However, in the presence of a tidal field, the isotropisation is much more rapid and pronounced, affecting the entire system rather than just the central region.

At first sight these results seem to indicate that the core formation and isotropisation seen in our 
simulations with tides are indeed tidally induced, and not related to the ROI. However, we cannot rule out that perhaps the tidal field triggers and/or accelerates the ROI. In particular, since the tides preferentially remove radial orbits, stripping will modify the angular momentum distribution of the system, which might well trigger the ROI. Potentially, then, both the isotropisation and core formation discussed here might be an outcome of the ROI, rather than merely a result of preferentially removing the more radial orbits. We emphasise that without a more detailed analysis, which is beyond the scope of this study, discriminating between these two explanations is non-trivial. In particlar, an obvious signature of the ROI is that it turns the spherical system into an elongated bar-like structure. However, tidal forces have the same effect. Indeed, all our stripped subhaloes, including the isotropic and tangentially anisotropic ones, have an elongated appearance at all times. To what extent this is an outcome of tides versus the ROI is something we hope to address in a forthcoming paper. For now, we leave it as an open question whether the core formation and isotropisation seen in our simulations are induced directly by the tides, or whether the tides trigger a ROI, which in turn transforms the central cusp into a core while driving the system towards isotropy.

\subsection{Breaking the mass\textendash anisotropy degeneracy of MW satellites}
\label{ssec:Breaking_Mass_Ani_Degeneracy}

Numerous studies in the past have used the radial velocities of individual stars in MW dwarf galaxies in order to constrain their mass profile \citep[e.g.][]{Strigari2008Natur4541096S, Walker2009ApJ7041274W, Chen2017MNRAS4681338C}. In particular, the central slope of the inferred dark matter halo is of crucial importance for potentially constraining the nature of dark matter \citep[e.g.][]{Zavala2019PhRvD100f3007Z, Fitts2019MNRAS490962F, SanchezAlmeida2024ApJ973L15S}. However, all these efforts are severely hampered by the well-known mass-anisotropy degeneracy \citep[e.g.][]{Binney.Mamon.82, Wilkinson2002MNRAS330778W, Read2017MNRAS4714541R}. As our analysis has shown, though, subhaloes experience efficient tidal isotropisation. Particularly for radially biased subhaloes, any pre-infall anisotropy is `wiped out' to leading order within $\sim 0.5 \torb$ (i.e., by the time of first peri-centric passage; see \fref{fig:Mass_Weighted_beta}). If this also affects the stellar component of the satellite galaxies that reside within these subhaloes, we have good reasons to assume that the systems are nearly isotropic, thereby alleviating the degeneracy. 
\begin{figure}
	\includegraphics[width=0.99\linewidth]{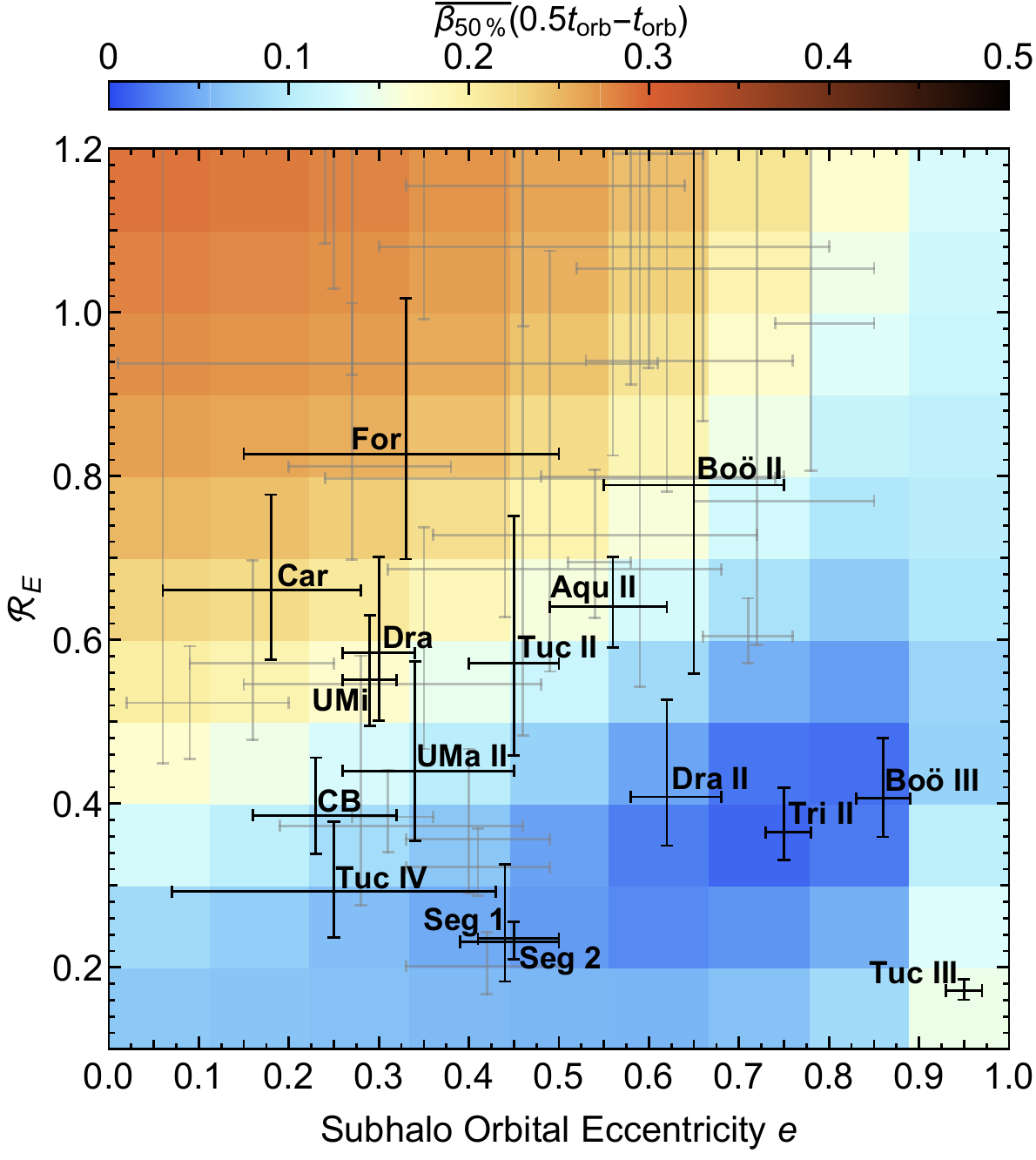}
	\caption{The orbital parameters ($\bigRE$ and $e$) of MW dwarf spheroidal satellite galaxies, taken from \citet{Pace2022ApJ940136P} (errorbars, indicating the 1$\sigma$ uncertainties). A subset of the more well-known dwarf galaxies is indicated in black and labelled, while the others are in grey. The background colour-shading indicates the average velocity anisotropy, $\overline{\beta_{50\%}}$, time-averaged over $0.5\torb$\textendash$\torb$, for subhaloes with constant $\beta = +0.5$ velocity anisotropy pre-infall.  Tidal isotropisation works to erase any pre-infall anisotropy $\beta(t)\rightarrow 0$ across the entire orbital parameter space. For sufficiently small $\bigRE$ and/or large $e$, this implies that the satellites may be expected to have close-to-isotropic velocity distributions. This alleviates the mass-anisotropy degeneracy that hampers a unique inference regarding their density profile from the observed line-of-sight velocities of their stars.}
	\label{fig:beta_Contour_plot}
\end{figure}

In order to better quantify this effect, we run a suite of simulations that tidally evolve the $\beta=+0.5$ subhalo on orbits that densely sample the entire orbital parameter space, covering $0.1 \leq \bigRE \leq 1.2$ and $0 \leq e \leq 1$. The colour-shading in \fref{fig:beta_Contour_plot} indicates the velocity anisotropy $\overline{\beta_{50\%}}$, time-averaged over $0.5\torb$\textendash$\torb$. Overall, the effect of tidal isotropisation becomes more pronounced with increasing maximal tidal field strength (equivalent to decreasing peri-centric distance), which corresponds to $e \rightarrow 1$ and/or $\bigRE \rightarrow 0$), reducing the initial $\beta_{50\%}=0.5$ (black) down to anywhere between $\overline{\beta_{50\%}} \simeq 0.3$ (orange) and $0.0$ (blue) during first infall. The fact that the trend slightly reverses towards the bottom-right corner of the $e$\textendash$\bigRE$ parameter space is simply due to the fact that along the highly radial orbits the tidal perturbations become highly impulsive, resulting in mass loss less than that along a circular orbit of similar $\bigRE$.

The fact that all subhaloes experience rapid tidal isotropisation allows us to place conservative constraints on the individual velocity anisotropy of the MW dwarf spheroidal satellite galaxies, whose measured orbital parameters from \citealt{Pace2022ApJ940136P} are overlaid in \fref{fig:beta_Contour_plot} as gray and black error bars.  Since subhaloes only become more and more isotropic over time, and many MW satellite galaxies have been orbiting the MW halo for more than a single orbit \citep{Fillingham2019arXiv190604180F, Battaglia2022AA657A}, the $\overline{\beta_{50\%}}(e, \bigRE)$ shown in Fig.~\ref{fig:beta_Contour_plot} may be considered a conservative upper limit for their amount of radial anisotropy. In the case of  Segue~1, Segue~2, Coma Berenices, Tucana~IV, Draco~II, Boötes~II and Triangulum~II we infer that $\overline{\beta_{50\%}} < 0.1$. Hence,  we may safely assume that at present these systems should be close to isotropic, removing any uncertainties regarding their inferred density profiles that result from the mass-anisotropy degeneracy. We emphasise, though, that we have assumed that the stellar body of a satellite galaxy undergoes the same level of tidal isotropisation as the subhalo it inhabits. To what extent this is accurate will need to be tested using idealised simulations of two-component anisotropic halo$+$star systems, which we leave for future work. 
 

\section{Summary and Conclusions}
\label{sec:Conclusions}

Although numerous papers in the past have used idealised simulations to study the tidal evolution of dark matter substructure (see \S\ref{sec:introduction}), without exception they have all focused on subhaloes that at infall are isotropic, i.e., have an ergodic distribution function $f=f(E)$. However, there is no a priori reason to assume that this is the case, and cosmological simulations have actually shown that dark matter haloes are predominantly radially anisotropic. In this paper we have used high-resolution, idealised simulations to study the impact of pre-infall velocity anisotropy on the tidal evolution of dark matter subhaloes. More specifically, we used the new Python package \texttt{PIANISTpy} to construct subhaloes with identical density distributions (truncated NFW haloes with concentration parameter $c=10$) that differ only in their initial velocity anisotropy, and evolved them along a wide range of orbits in an analytical, static host halo potential using the AMR code $\gamer$. 

We find that the tidal evolution of anisotropic subhaloes differs substantially from that of their isotropic counterparts. In particular:
\begin{itemize}
	
\item Radially (tangentially) anisotropic subhaloes experience more (less) tidal mass loss than their isotropic counterparts, irrespective of the subhalo's orbital parameters. Quantitatively, the bound mass fraction of an anisotropic subhalo can already differ from that of the conventional, isotropic predictions by order unity after a single orbit and by several orders of magnitude after a Hubble time. This strong dependence of the tidal evolution on the internal orbital anisotropy is a consequence of the fact that more radial orbits have larger apo-centres compared to more tangential orbits of the same orbital energy. Since stripping mainly removes orbits that reach outside of the tidal radius, more radial orbits are therefore more easily stripped. 

\item The strong dependence on the internal velocity anisotropy implies that the tidal tracks that are often used in (semi)-analytical models of tidal stripping are not universal. In other words, knowing the density profile of a subhalo at infall is not sufficient to predict its tidal evolution. However, systems with the same density {\it and} the same anisotropy profiles {\it do} evolve along unique tidal tracks, such that their structure at later times depends only on its instantaneous bound mass fraction, independent of the orbit along which it evolves. 

\item Subhaloes that are radially anisotropic throughout with $\beta \gta 0.3$ can undergo a tidally induced cusp-to-core transformation. This is a consequence of the fact that highly radial orbits that contribute density to the central cusp can still have apo-centres that fall outside of the tidal radius. Hence, their stripping directly lowers the central density, which, aided by the re-virialisation that follows the removal of mass, can result in rapid core formation. Once the core radius becomes comparable to the instantaneous tidal radius, these systems can undergo complete, physical tidal disruption \citep[see also][]{Penarrubia2010MNRAS4061290P}. Isotropic and tangentially anisotropic subhaloes, as well as Osipkov\textendash Merrit systems with an isotropic cusp, do not undergo tidal core formation, and are therefore immune to tidal disruption \citep[e.g.,][]{vandenBosch2018MNRAS4743043V, Errani2020MNRAS491, Errani2021MNRAS50518E, Amorisco2021arXiv211101148A, Stucker2023MNRAS5214432S}.

\item Because of the preferential stripping of more radial orbits, subhaloes that at infall are radially anisotropic become more and more isotropic as time goes on. In addition, along eccentric orbits, the tidal impulsive shocks associated with peri-centric passages cause a mixing in energy-angular momentum space that drives the velocity distribution of the bound remnant towards isotropy, even for systems that are tangentially anisotropic at infall. Hence, in general, as more and more mass is stripped off a subhalo, the remnant becomes more and more isotropic, a process we term `tidal isotropisation'. In the case of subhaloes that are radially anisotropic at infall, the isotropisation is extremely fast, with the entire system having become isotropic within less than one orbital time. Subhaloes that start out tangentially anisotropic evolve less rapidly, and less strongly, mainly driven by the impulsive, tidal shocks at peri-centric passages.

\end{itemize}

These findings have several important implications. First of all, cosmological simulations have revealed a significant halo-to-halo variance \citep[e.g.][]{Ludlow2011MNRAS4153895L, Klypin2016MNRAS4574340K, Cataldi2021MNRAS5015679C} and strong halo-mass-dependence \citep{He2024arXiv240714827H} in the velocity anisotropy profiles of dark matter haloes. Consequently, given that the tidal tracks depend strongly on the velocity anisotropy at infall, the structural properties of tidally evolved subhaloes are expected to be more diverse than typically modeled, displaying rather significant scatter in their correlations between structural parameters (i.e.,  $V_{\rm max}$ and $r_{\rm max}$) and the bound mass fraction. 

Secondly, the tidal core formation process highlighted here is an important alternative to other core formation processes such as dynamical friction \citep[][]{elzant01, Goerdt.etal.10} and episodic SN-driven outflows \citep[][]{Pontzen.Governato.12, Teyssier.etal.13, elzant16}. Most importantly, the latter is only expected to be effective in dark matter haloes spanning the range $10^{10} h^{-1}M_\odot \lta M_\rmh \lta 10^{11} h^{-1}M_\odot$ \citep[][]{DiCintio.etal.14, Tollet.etal.16, Freundlich.etal.20b}. Although AGN-driven outflows may be relevant for creating cores in more massive haloes \citep[e.g.,][]{Martizzi.etal.13, Dekel.etal.21}, baryonic processes are not expected to be able to create cores in low mass haloes with $M_\rmh \lta 10^{10} h^{-1}M_\odot$, simply because the stellar masses of the galaxies hosted in these haloes are too low for SN feedback to be impactful.  For this reason, it is generally argued that, in the CDM scenario, the (sub)haloes hosting ultra-faint dwarf galaxies in the MW should be cusped \citep[][]{Bullock.BoylanKolchin.17}. However, the tidal core-formation process highlighted here operates on {\it all} mass scales, as long as the halo at infall was sufficiently radially anisotropic. Hence, if the subhaloes hosting ultra-faint dwarf galaxies are found to be cored, this would not necessarily rule against CDM.

Finally, tidal isotropisation works to erase any pre-infall subhalo velocity anisotropy. Under the assumption that the satellite galaxies that reside in subhaloes also undergo tidal isotropisation, this predicts that satellite galaxies that were accreted sufficiently long ago should have a close-to-isotropic velocity distribution. This helps to break the mass-anisotropy degeneracy that hampers accurate constraint on the mass profiles of satellite galaxies from modeling the line-of-sight velocities of their stars. Applying this idea to MW dwarf spheroidals with reliable constraints on their orbital parameters, we infer that Segue~1, Segue~2, Coma Berenices, Tucana~IV, Draco~II, Boötes~II and Triangulum~II are unlikely to have any significant anisotropy, and are thus ideally suited for dynamical studies of their mass profiles.

\section*{Acknowledgements}

We acknowledge useful conversations with Rapha\"el Errani, Marla Geha and Go Ogiya. We also thank the anonymous referee for a constructive and detailed report. FvdB is supported by the National Science Foundation (NSF) through grants AST-2307280 and AST-2407063. This research is partially supported by the National Science and Technology Council (NSTC) of Taiwan under Grant No. NSTC 111-2628-M-002-005-MY4 and the National Taiwan University (NTU) Academic Research-Career Development Project under Grant No. NTU-CDP-113L7729. We use \texttt{NumPy} \citep{numpy} and \texttt{SciPy} \citep{scipy} for data analysis, and \texttt{Matplotlib} \citep{matplotlib} and \texttt{yt} \citep{yt} for data visualisation.

\section*{Data Availability}

The data underlying this article will be shared on reasonable request to the corresponding author.

\appendix

\section{Numerical Convergence and Stability Tests}
\label{app:Numerical_Convergence}

Here we present various numerical convergence tests to gauge the robustness of the simulations carried out as part of this study.  Appendix~\ref{app:Isolation} demonstrates that, except for the radial orbit instability that affects the systems with large, initial radial anisotropy, our initial conditions are stable and that numerical heating is completely negligible. Appendix~\ref{app:treecode_GAMER} compares the performance of $\gamer$ with that of a {\tt treecode}.
\begin{figure*}
	\includegraphics[width=\linewidth]{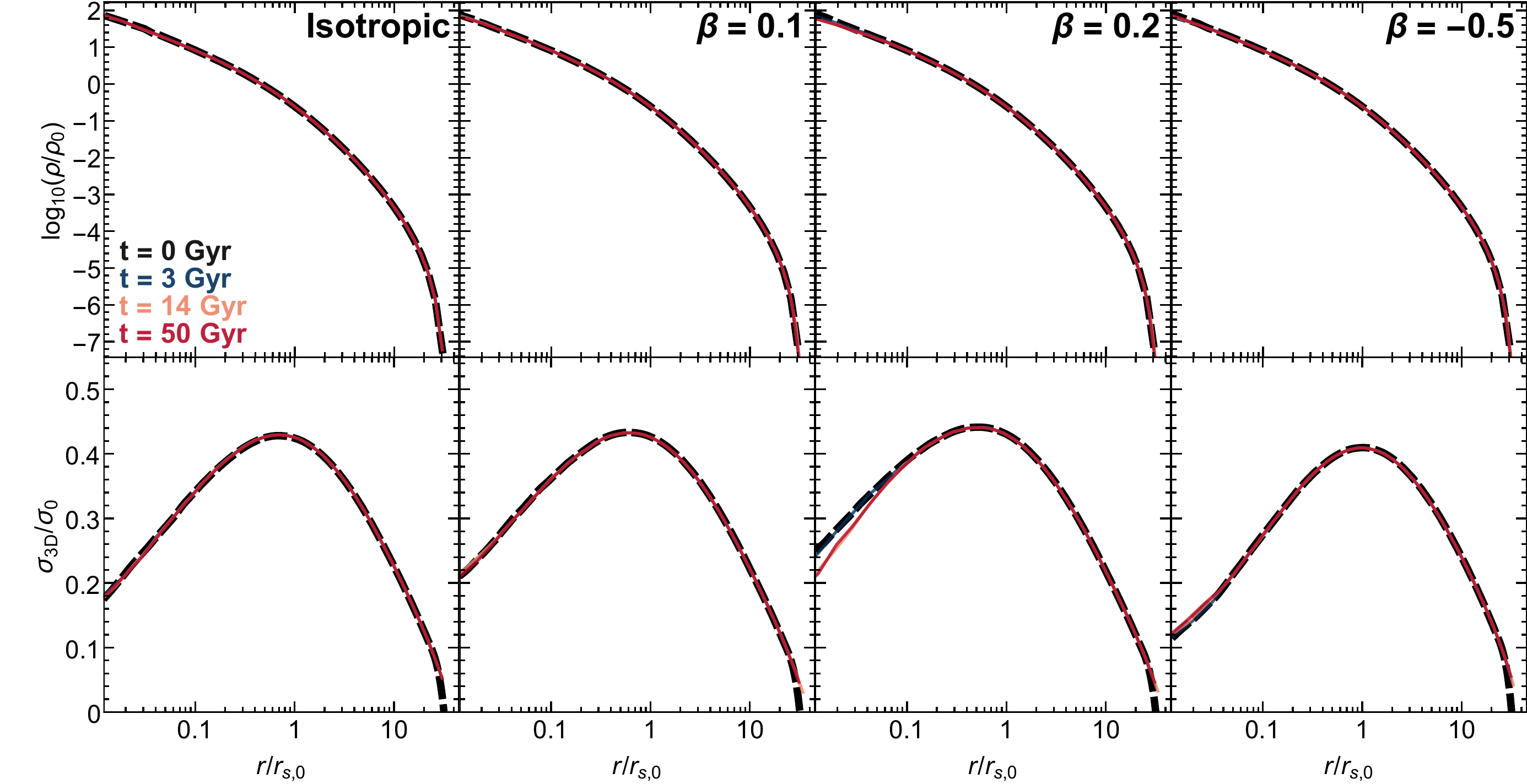}
	\caption{Time evolution of shell-averaged density (top row) and 3D velocity dispersion profiles (bottom) of the isotropic, $\beta = 0.1, 0.2,$ and $-0.5$ subhaloes (leftmost to rightmost columns) in isolation for 50~Gyr~$\simeq 25 \tcross$ with the AMR and simulation setup described in \S\ref{ssec:numsim}. Artificial two-body heating remains negligible down to grid resolution scale $r = 0.012 \rsz$, demonstrating both the stability of \texttt{PIANISTpy}-constructed phase-space-truncated initial conditions and numerical accuracy of $\gamer$ over the resolved range of length scales.}
	\label{fig:Isolated_NFW_Evolution}
\end{figure*}

\subsection{Numerical stability and heating in isolation}
\label{app:Isolation}

To verify that our AMR refinement strategy (see \S\ref{ssec:numsim}) is adequate, and that our initial conditions are stable, we dynamically evolve the isotropic, $\beta = +0.2, +0.1,$ and $-0.5$ haloes, generated with $\Npar = 10^7$ equal-mass particles using \texttt{PIANISTpy}, in isolation for $50\Gyr \simeq 25 \tcross$. \fref{fig:Isolated_NFW_Evolution} shows the resulting density (top row) and 3D velocity dispersion profiles (bottom) at four different epochs: $t = 0$ (black dashed), 3 (blue), 14 (orange), and $50$~Gyr (red). Different columns correspond to haloes with a different initial anisotropy, as indicated. Evidently, the isotropic and $\beta = +0.1$ and $-0.5$ systems are extremely stable, showing no sign of evolution from the initial conditions for the full duration of each simulation. Hence, we see that, $\gamer$, with the ARM refinement strategy adopted here, provides sufficient force and mass resolutions, free of any discernible numerical two-body relaxation effect. In the case of $\beta=+0.2$, some late-time evolution in the 3D velocity dispersion is apparent inside of $0.1 \rsz$. As discussed in detail in Chiang et al. (in preparation), this owes to the onset of the ROI (cf. Fig.~\ref{fig:beta_Evolution_Cir}). 

\subsection{Code comparison}
\label{app:treecode_GAMER}

As this is the first time that $\gamer$ is used to study the tidal evolution of substructure, it is judicious to compare its performance with a more tested code. We therefore aim to reproduce some results of \citet{vdBosch.Ogiya.18}, who used the hierarchical $N$-body code {\tt treecode}, written by J. Barnes and with some improvements due to J. Dubinski, to evolve isotropic $N$-body subhaloes in the analytical potential of a host halo. {\tt treecode} uses a \citet{Barnes.Hut.86} octree to compute accelerations based on a multipole expansion up to quadrupole order, and uses a second-order leap-frog integration scheme to solve the equations of motion. Forces between particles are softened using a simple Plummer softening.
\begin{figure}
	\includegraphics[width=\linewidth]{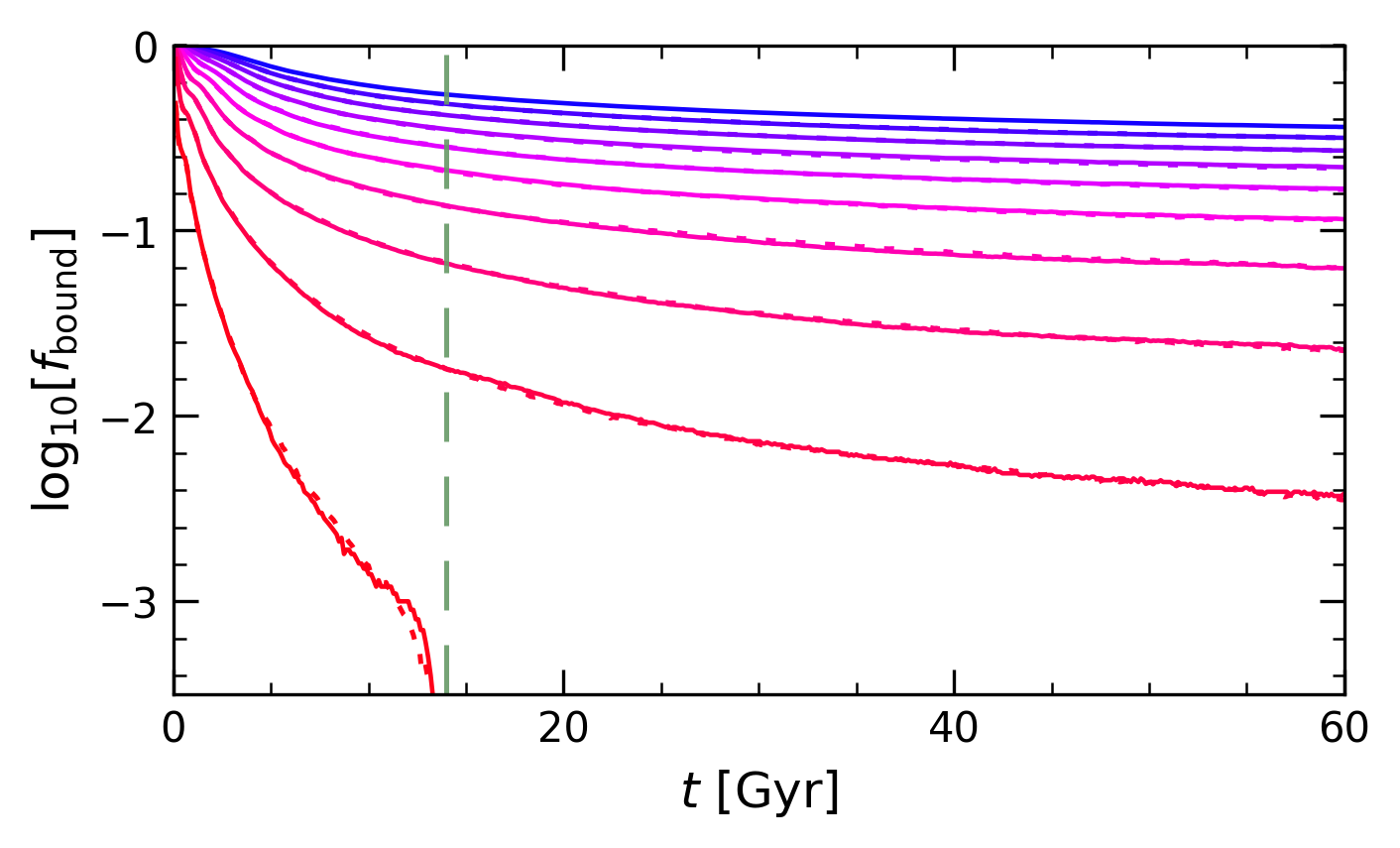}
	\caption{Time evolution of the bound mass fractions $\fbound$ for hard-truncated isotropic subhaloes resolved with $\Npar = 10^5$ on circular orbits with $\bigRE = 1.0, 0.9, ..., 0.1$ (blue to red).
    The dashed curves show the results obtained here using $\gamer$, and are compared to the results of 
    \citet{vdBosch.Ogiya.18} obtained using \texttt{treecode} (solid lines, taken from their Fig.~4). The vertical dashed line marks the Hubble time $\tH = 13.97$~Gyr. The relative differences are below $|\Delta\fbound/\fbound| \leq 5\%$ across all ten cases.}
	\label{fig:f_bound_treecode_GAMER}
\end{figure}

We use $\gamer$ to evolve the initial condition files of \citet{vdBosch.Ogiya.18} (hard-truncated, isotropic NFW subhaloes with $\Npar = 10^5$) along circular orbits in the same analytical host halo as used in their study. \fref{fig:f_bound_treecode_GAMER} compares the resulting bound mass fractions as function of time (dashed lines) to those of  \citet{vdBosch.Ogiya.18} (solid lines, taken from their Fig.~4). Different colours correspond to different circular orbits, with $\bigRE$ ranging from $1.0$ (blue) to $0.1$ (red) in steps of $0.1$. Note the excellent agreement between the $\gamer$ and {\tt treecode} results, with relative differences smaller than $|\Delta\fbound(t)/\fbound(t)| \leq 5\%$ across all ten cases simulated. This validates  $\gamer$ as a suitable and competitive $N$-body code.



\bibliographystyle{mnras}
\bibliography{MyBibTeX1} 

\label{lastpage}
\end{document}